\documentclass[12pt,prd,showpacs,tightenlines,nofootinbib]{revtex4}
\usepackage{bm}
\usepackage{graphics}
\usepackage{rotating}
\usepackage{epsfig}
\begin{document}
\title{Weak decays of $B_s$ mesons to $D_s$ mesons in the
  relativistic quark model } 

\author{R. N. Faustov}
\author{V. O. Galkin}
\affiliation{Dorodnicyn Computing Centre, Russian Academy of Sciences,
  Vavilov Str. 40, 119333 Moscow, Russia}

\begin{abstract}
The form factors of weak decays of $B_s$ mesons to ground state $D_s^{(*)}$
mesons as well as to their radial $D_s^{(*)}(2S)$ and orbital
$D_{sJ}^{(*)}$ excitations are calculated in the framework of the
relativistic quark model based on the quasipotential approach.
All relativistic effects, including
contributions of intermediate negative-energy states and boosts of the
meson wave functions, are consistently taken into account. As a result
the form factors are determined in the whole kinematical range without
additional phenomenological parametrizations and extrapolations. On
this basis semileptonic decay branching fractions are
calculated. Two-body nonleptonic $B_s$ decays are considered within the
factorization approximation. The obtained results agree well with
available experimental data.  

\end{abstract}

\pacs{ 13.20.He, 12.39.Ki}

\maketitle

\section{Introduction}
\label{sec:int}

In recent years significant experimental progress has been achieved
in studying properties of $B_s$ mesons \cite{pdg}. The Belle Collaboration
considerably increased the number of observed $B_s$ mesons and their
decays due to the data collected in $e^+e^-$ collisions at the
$\Upsilon(10860)$ resonance \cite{belle1}. On the other hand, $B_s$
mesons are copiously produced at Large Hadron Collider
(LHC). First precise data on their properties are coming from the LHCb
Collaboration \cite{lhcb1}. Several weak decay modes of the $B_s$
meson were observed for the first time \cite{lhcb1p}.  New data are expected in
near future  \cite{lhcb2}. The study of weak $B_s$ decays  is
important for further improvement in the determination of the
Cabibbo-Kobayashi-Maskawa (CKM) matrix elements, for testing the
prediction  of the Standard Model and searching for possible
deviations from these predictions, the so-called ``new
physics''.  

The dominant decay channel of the $B_s$ meson is into the $D_s$ meson plus
anything \cite{pdg}. Therefore various important properties of excited
$D_s$ mesons can be studied in the $B_s$ meson weak decays. In particular, they can
shed light on the controversial $D^*_{s0}(2317)$ and $D_{s1}(2460)$ mesons, whose
nature still remains unclear in the literature. The abnormally light
masses of these mesons put them below $DK$ and $D^*K$ thresholds, thus
making these states narrow since the only allowed  decays
violate isospin. The peculiar feature of these mesons is that they
have masses almost equal to or even lower than the masses of their
charmed counterparts  $D^*_0(2400)$ and $D_1(2420)$ \cite{pdg}. If
these mesons are indeed $1^3P_0$ and $1P_1$ states, then all $1P$
states of the $D_s$ meson are narrow, contrary to the $D$ meson
case. This narrowness considerably simplifies the experimental
investigation of weak $B_s$ 
decays to orbitally excited $D_{sJ}^{(*)}$ mesons.        
Recently it was proposed \cite{blorrs} that study of $B_s\to D_{sJ}^{(*)}$ transitions can
clarify some puzzles in the corresponding semileptonic $B$ decays. 

In this paper, we  extend our investigations of weak $B$ and $B_c$ decays
\cite{hlsem,bcexc} to studying exclusive weak semileptonic and nonleptonic 
decays of the $B_s$ to the ground state, radially and orbitally excited
$D_s$ mesons.  For the calculations we use the same effective methods
\cite{hlsem,bcexc} previously developed and successfully applied in the framework 
of the QCD-motivated relativistic quark model based on the quasipotential
approach.   The weak decay matrix elements are parametrized by the
invariant form factors which are then expressed through the overlap integrals of
the meson wave functions. The systematic account for relativistic
effects, including the wave function transformations to the
moving reference frame and  contributions from the intermediate negative-energy
states, allows one to reliably determine the momentum transfer dependence
of the decay form factors in the whole accessible kinematical range. The
other important advantage of this approach is that for numerical calculations we use
the relativistic wave functions, obtained in the
meson mass spectra calculations \cite{mass,hlm}. Thus we do not need 
any additional ad hoc parametrizations or extrapolations
which were usually used in some previous investigations. 

The calculated
form factors are then substituted in expressions for the differential
decay rates and semileptonic decay branching fractions are
evaluated. The tree-dominated two-body nonleptonic
$B_s$ decays to the $D_s$ meson and light or another $D_s$ meson are
studied on the basis of the factorization  approach. Such approximation
significantly simplifies calculations, since it allows one to
express the matrix elements of the weak 
Hamiltonian governing  the nonleptonic decays through the product of
the transition matrix elements and meson weak decay constants. All these
ingredients are available in our model. The obtained results are
compared with previous calculations and experimental values, which are
measured for some $B_s$ semi-exclusive semileptonic and several
exclusive nonleptonic decay modes. 

The paper is organized as follows.   In Sec.~\ref{rqm} we briefly
describe the relativistic quark model. Then in Sec.~\ref{mml}   we
discuss the relativistic calculation of the transition matrix element of
the weak $b\to c$ current between meson states in the quasipotential
approach. Special attention is paid to the contributions of the
negative energy states and the relativistic transformation of the wave
functions to the moving reference frame. These methods
are applied in Sec.~\ref{sec:ffsdgs} to the calculation of the form factors of weak $B_s$
decays to ground state $D_s$ mesons. The form
factors are obtained as the overlap integrals of meson wave functions
within the heavy quark expansion up to subleading order. It is shown that
all heavy quark symmetry relations are explicitly satisfied. 
These form factors are used for evaluating semileptonic decay
branching fractions in Sec.~\ref{sdgs}. The calculations of the form factors and
semileptonic decay branching 
fractions for $B_s$ decays to radially excited $D_s^{(*)}(2S)$ mesons are
presented in Secs.~\ref{sec:ffrexc} and \ref{sdre} within the same
approach. In Sec.~\ref{sec:fforbexc} the form
factors of weak $B_s$ decays to orbitally excited $D_{sJ}^{(*)}$
mesons are obtained. Semileptonic branching fractions
for $B_s$ decays to orbitally excited $D_{sJ}^{(*)}$ mesons are given in
Sec.~\ref{sdoe}. Finally, the two-body nonleptonic $B_s$ decays
calculated within the factorization approximation are presented in
Sec.~\ref{nl}. All obtained results are confronted with previous
calculations and available experimental data. Section~\ref{sec:concl}
contains the conclusions. The relations between the sets of weak form factors, the
model independent HQET expressions for the form factors, and helicity
components of the hadronic tensor defined in terms of the form factors are presented
in the Appendices.

\section{Relativistic quark model}  
\label{rqm}

In the quasipotential approach a meson is described as a bound
quark-antiquark state with a wave function satisfying the
quasipotential equation of the Schr\"odinger type \cite{mass}
\begin{equation}
\label{quas}
{\left(\frac{b^2(M)}{2\mu_{R}}-\frac{{\bf
p}^2}{2\mu_{R}}\right)\Psi_{M}({\bf p})} =\int\frac{d^3 q}{(2\pi)^3}
 V({\bf p,q};M)\Psi_{M}({\bf q}),
\end{equation}
where the relativistic reduced mass is
\begin{equation}
\mu_{R}=\frac{E_1E_2}{E_1+E_2}=\frac{M^4-(m^2_1-m^2_2)^2}{4M^3},
\end{equation}
and $E_1$, $E_2$ are the center of mass energies on mass shell given by
\begin{equation}
\label{ee}
E_1=\frac{M^2-m_2^2+m_1^2}{2M}, \quad E_2=\frac{M^2-m_1^2+m_2^2}{2M}.
\end{equation}
Here $M=E_1+E_2$ is the meson mass, $m_{1,2}$ are the quark masses,
and ${\bf p}$ is their relative momentum.  
In the center of mass system the relative momentum squared on mass shell 
reads
\begin{equation}
{b^2(M) }
=\frac{[M^2-(m_1+m_2)^2][M^2-(m_1-m_2)^2]}{4M^2}.
\end{equation}

The kernel 
$V({\bf p,q};M)$ in Eq.~(\ref{quas}) is the quasipotential operator of
the quark-antiquark interaction. It is constructed with the help of the
off-mass-shell scattering amplitude, projected onto the positive
energy states. 
Constructing the quasipotential of the quark-antiquark interaction, 
we have assumed that the effective
interaction is the sum of the usual one-gluon exchange term with the mixture
of long-range vector and scalar linear confining potentials, where
the vector confining potential
contains the Pauli interaction. The quasipotential is then defined by
\cite{mass}
  \begin{equation}
\label{qpot}
V({\bf p,q};M)=\bar{u}_1(p)\bar{u}_2(-p){\mathcal V}({\bf p}, {\bf
q};M)u_1(q)u_2(-q),
\end{equation}
with
$${\mathcal V}({\bf p},{\bf q};M)=\frac{4}{3}\alpha_sD_{ \mu\nu}({\bf
k})\gamma_1^{\mu}\gamma_2^{\nu}
+V^V_{\rm conf}({\bf k})\Gamma_1^{\mu}
\Gamma_{2;\mu}+V^S_{\rm conf}({\bf k}),$$
where $\alpha_s$ is the QCD coupling constant, $D_{\mu\nu}$ is the
gluon propagator in the Coulomb gauge
\begin{equation}
D^{00}({\bf k})=-\frac{4\pi}{{\bf k}^2}, \quad D^{ij}({\bf k})=
-\frac{4\pi}{k^2}\left(\delta^{ij}-\frac{k^ik^j}{{\bf k}^2}\right),
\quad D^{0i}=D^{i0}=0,
\end{equation}
and ${\bf k=p-q}$. Here $\gamma_{\mu}$ and $u(p)$ are 
the Dirac matrices and spinors
\begin{equation}
\label{spinor}
u^\lambda({p})=\sqrt{\frac{\epsilon(p)+m}{2\epsilon(p)}}
\left(
\begin{array}{c}1\cr {\displaystyle\frac{\bm{\sigma}
      {\bf  p}}{\epsilon(p)+m}}
\end{array}\right)\chi^\lambda,
\end{equation}
where  $\bm{\sigma}$   and $\chi^\lambda$
are Pauli matrices and spinors, respectively, and $\epsilon(p)=\sqrt{{\bf p}^2+m^2}$.
The effective long-range vector vertex is
given by
\begin{equation}
\label{kappa}
\Gamma_{\mu}({\bf k})=\gamma_{\mu}+
\frac{i\kappa}{2m}\sigma_{\mu\nu}k^{\nu},
\end{equation}
where $\kappa$ is the Pauli interaction constant characterizing the
long-range anomalous chromomagnetic moment of quarks. Vector and
scalar confining potentials in the nonrelativistic limit reduce to
\begin{eqnarray}
\label{vlin}
V_{\rm conf}^V(r)&=&(1-\varepsilon)(Ar+B),\nonumber\\ 
V_{\rm conf}^S(r)& =&\varepsilon (Ar+B),
\end{eqnarray}
reproducing 
\begin{equation}
\label{nr}
V_{\rm conf}(r)=V_{\rm conf}^S(r)+V_{\rm conf}^V(r)=Ar+B,
\end{equation}
where $\varepsilon$ is the mixing coefficient. 

The expression for the quasipotential of the heavy quarkonia
within and without the $v^2/c^2$ expansion  can be found in Ref.~\cite{mass}. The
quasipotential for the heavy quark interaction with a light antiquark
without employing the nonrelativistic ($v/c$)  expansion 
is given in Ref.~\cite{hlm}.  All the parameters of
our model like quark masses, parameters of the linear confining potential
$A$ and $B$, mixing coefficient $\varepsilon$ and anomalous
chromomagnetic quark moment $\kappa$ are fixed from the analysis of
heavy quarkonium masses and radiative
decays  \cite{mass}. The quark masses
$m_b=4.88$ GeV, $m_c=1.55$ GeV, $m_s=0.5$ GeV, $m_{u,d}=0.33$ GeV and
the parameters of the linear potential $A=0.18$ GeV$^2$ and $B=-0.30$ GeV
have values inherent for quark models.  The value of the mixing
coefficient of vector and scalar confining potentials $\varepsilon=-1$
has been determined from the consideration of the heavy quark expansion
for the semileptonic $B\to D$ decays
\cite{fg} and charmonium radiative decays \cite{mass}.
Finally, the universal Pauli interaction constant $\kappa=-1$ has been
fixed from the analysis of the fine splitting of heavy quarkonia ${
}^3P_J$- states \cite{mass} and  the heavy quark expansion for semileptonic
decays of heavy mesons \cite{fg} and baryons \cite{sbar}. Note that the 
long-range  magnetic contribution to the potential in our model
is proportional to $(1+\kappa)$ and thus vanishes for the 
chosen value of $\kappa=-1$ in accordance with the flux tube model.

\section{Matrix elements of the electroweak current for
  ${\lowercase{b\to c}}$ transition} \label{mml}

In order to calculate the exclusive semileptonic decay rate of the
$B_s$ meson, it is necessary to determine the corresponding matrix
element of the  weak current between meson states.
In the quasipotential approach,  the matrix element of the weak current
$J^W_\mu=\bar c\gamma_\mu(1-\gamma_5)b$, associated with the $b\to c$  transition, between a $B_s$ meson with mass $M_{B_s}$ and
momentum $p_{B_s}$ and a final $D_s$ meson with mass $M_{D_s}$ and momentum $p_{D_s}$ takes the form \cite{f}
\begin{equation}\label{mxet} 
\langle D_s(p_{D_s}) \vert J^W_\mu \vert B_s(p_{B_s})\rangle
=\int \frac{d^3p\, d^3q}{(2\pi )^6} \bar \Psi_{{D_s}\,{\bf p}_{D_s}}({\bf
p})\Gamma _\mu ({\bf p},{\bf q})\Psi_{B_s\,{\bf p}_{B_s}}({\bf q}),
\end{equation}
where $\Gamma _\mu ({\bf p},{\bf
q})$ is the two-particle vertex function and  
$\Psi_{M\,{\bf p}_M}$ are the
meson ($M=B_s,{D_s})$ wave functions projected onto the positive energy states of
quarks and boosted to the moving reference frame with momentum ${\bf p}_M$.
\begin{figure}
  \centering
  \includegraphics{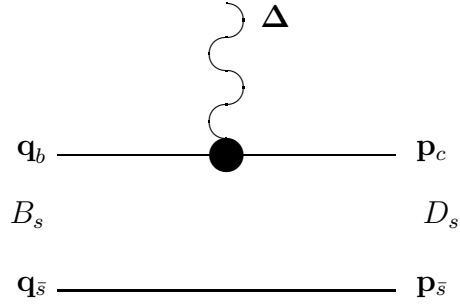}
\caption{Lowest order vertex function $\Gamma^{(1)}({\bf p},{\bf q})$
contributing to the current matrix element (\ref{mxet}). \label{d1}}
\end{figure}

\begin{figure}
  \centering
  \includegraphics{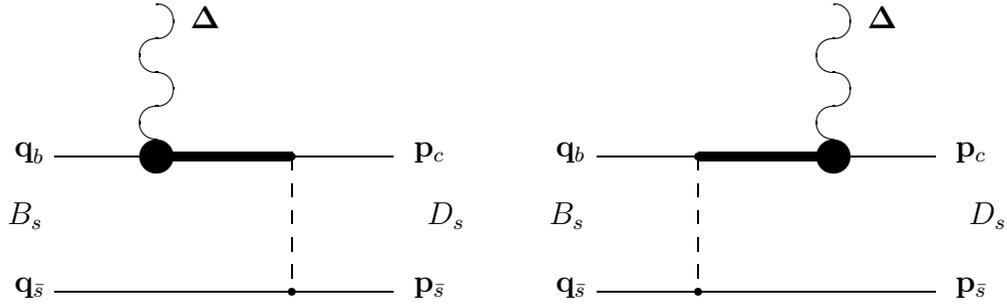}
\caption{ Vertex function $\Gamma^{(2)}({\bf p},{\bf q})$
taking the quark interaction into account. Dashed lines correspond  
to the effective potential ${\cal V}$ in 
(\ref{qpot}). Bold lines denote the negative-energy part of the quark
propagator. \label{d2}}
\end{figure}

 The contributions to $\Gamma_\mu ({\bf p},{\bf q})$ come from Figs.~\ref{d1} and \ref{d2}. 
The contribution $\Gamma^{(2)}_\mu ({\bf p},{\bf q})$ is the consequence
of the projection onto the positive-energy states. Note that the form of the
relativistic corrections emerging from the vertex function
$\Gamma^{(2)}_\mu ({\bf p},{\bf q})$ explicitly depends on the Lorentz structure of the
quark-antiquark interaction. In the heavy quark limit $m_{Q}\to \infty$, 
only $\Gamma^{(1)}_\mu ({\bf p},{\bf q})$ contributes, while $\Gamma^{(2)}_\mu ({\bf p},{\bf q})$  
give contributions starting from the subleading order. 
The vertex functions look like
\begin{equation} \label{gamma1}
\Gamma_\mu^{(1)}({\bf
p},{\bf q})=\bar u_{c}(p_c)\gamma_\mu(1-\gamma^5)u_b(q_b)
(2\pi)^3\delta({\bf p}_{\bar s}-{\bf
q}_{\bar s}),\end{equation}
and
\begin{eqnarray}\label{gamma2} 
\Gamma_\mu^{(2)}({\bf
p},{\bf q})&=&\bar u_{c}(p_c)\bar u_s(p_{\bar s}) \Bigl\{{\cal V}({\bf p}_{\bar s}-{\bf
q}_{\bar s})\frac{\Lambda_{c}^{(-)}(k')}{ \epsilon_{c}(k')+
\epsilon_{c}(q_b)}\gamma_1^0 \gamma_{1\mu}(1-\gamma_1^5)\nonumber \\ 
& &+\gamma_{1\mu}(1-\gamma_1^5)
\frac{\Lambda_b^{(-)}(
k)}{\epsilon_b(k)+\epsilon_b(p_c)}\gamma_1^0
{\cal V}({\bf p}_{\bar s}-{\bf
q}_{\bar s})\Bigr\}u_b(q_b)
u_s(q_{\bar s}),\end{eqnarray}
where the superscripts ``(1)" and ``(2)" correspond to Figs.~\ref{d1} and
\ref{d2},  ${\bf k}={\bf p}_c-{\bf\Delta};\
{\bf k}'={\bf q}_b+{\bf\Delta};\ {\bf\Delta}={\bf
p}_{D_s}-{\bf p}_{B_s}$;
$$\Lambda^{(-)}(p)=\frac{\epsilon(p)-\bigl( m\gamma
^0+\gamma^0({\bm{ \gamma}{\bf p}})\bigr)}{ 2\epsilon (p)}.$$
Here \cite{f} 
\begin{eqnarray*} 
p_{c,\bar s}&=&\epsilon_{c,s}(p)\frac{p_{D_s}}{M_{D_s}}
\pm\sum_{i=1}^3 n^{(i)}(p_{D_s})p^i,\\
q_{b,\bar s}&=&\epsilon_{b,s}(q)\frac{p_{B_s}}{M_{B_s}} \pm \sum_{i=1}^3 n^{(i)}
(p_{B_s})q^i,\end{eqnarray*}
and $n^{(i)}$ are three four-vectors given by
$$ n^{(i)\mu}(p)=\left\{ \frac{p^i}{M},\ \delta_{ij}+
\frac{p^ip^j}{M(E+M)}\right\}, \quad E=\sqrt{{\bf p}^2+M^2}.$$

The wave function of a final $D_{s}$ meson at rest is given by
\begin{equation}\label{psi}
\Psi_{{D_s}}({\bf p})\equiv
\Psi^{JLS{\cal M}}_{D_{sJ}}({\bf p})={\cal Y}^{JLS{\cal M}}\,\psi_{D_{sJ}}({\bf p}),
\end{equation}
where $J$ and ${\cal M}$ are the total meson angular momentum and its
projection,  $L$ is the orbital momentum,
while $S=0,1$ is the total spin.   
$\psi_{D_{sJ}}({\bf p})$ is the radial part of the wave function,
which has been determined by the numerical solution of Eq.~(\ref{quas})
in Ref.~\cite{hlm}.
The spin-angular momentum part ${\cal Y}^{JLS{\cal M}}$ has the following form
\begin{equation}\label{angl}
{\cal Y}^{JLS{\cal M}}=\sum_{\sigma_1\sigma_2}\langle L\, {\cal M}-\sigma_1-\sigma_2,\  
S\, \sigma_1+\sigma_2 |J\, {\cal M}\rangle\langle \frac12\, \sigma_1,\ 
\frac12\, \sigma_2 |S\, \sigma_1+\sigma_2\rangle Y_{L}^{{\cal M}-\sigma_1-\sigma_2}
\chi_1(\sigma_1)\chi_2(\sigma_2).
\end{equation}
Here $\langle j_1\, m_1,\  j_2\, m_2|J\, {\cal M}\rangle$ are the Clebsch-Gordan 
coefficients, $Y_l^m$ are the spherical harmonics, and $\chi(\sigma)$ (where 
$\sigma=\pm 1/2$) are the spin wave functions,
$$ \chi\left(1/2\right)={1\choose 0}, \qquad 
\chi\left(-1/2\right)={0\choose 1}. $$

The heavy-light meson states ($D_{s1}$, $D_{s1}'$)   with $J=L=1$
are the mixtures of spin-triplet $D_s(^3P_1)$  and spin-singlet $D_s(^1P_1)$
states:
\begin{eqnarray}
  \label{eq:mix}
  \Psi_{D_{s1}}&=&\Psi_{D_s(^1P_1)}\cos\varphi+\Psi_{D_s(^3P_1)}\sin\varphi, \cr
 \Psi_{D_{s1}'}&=&-\Psi_{D_s(^1P_1)}\sin\varphi+\Psi_{D_s(^3P_1)}\cos\varphi, 
\end{eqnarray}
where $\varphi=34.5^{\circ}$ is the mixing angle and the primed state
has the heavier mass \cite{hlm}.
  Such mixing occurs due to the nondiagonal spin-orbit and
tensor terms in the $Q\bar q$ quasipotential. The physical states are obtained
by diagonalizing the corresponding mixing terms. Note that the above value
of the mixing angle $\varphi$ is very close to its heavy quark limit
$\varphi_{m_Q\to\infty}=\arctan(\sqrt{1/2})\approx35.3^{\circ}$. This
means that the wave functions $\Psi_{D_{s1}}$ and $ \Psi_{D'_{s1}}$
correspond in the heavy quark limit to
$\Psi_{D_{s(3/2)}}$ and $\Psi_{D_{s(1/2)}}$, respectively.

It is important to point out that the wave functions entering the weak current
matrix element (\ref{mxet}) are not in the rest frame in general. For example, 
in the $B_s$ meson rest frame (${\bf p}_{B_s}=0$), the final  meson
is moving with the recoil momentum ${\bf \Delta}$. The wave function
of the moving  meson $\Psi_{D_s\,{\bf\Delta}}$ is connected 
with the  wave function in the rest frame 
$\Psi_{D_s\,{\bf 0}}\equiv \Psi_{D_s}$ by the transformation \cite{f}
\begin{equation}
\label{wig}
\Psi_{D_s\,{\bf\Delta}}({\bf
p})=D_c^{1/2}(R_{L_{\bf\Delta}}^W)D_s^{1/2}(R_{L_{
\bf\Delta}}^W)\Psi_{D_s\,{\bf 0}}({\bf p}),
\end{equation}
where $R^W$ is the Wigner rotation, $L_{\bf\Delta}$ is the Lorentz boost
from the meson rest frame to a moving one, and   
the rotation matrix $D^{1/2}(R)$ in spinor representation is given by
\begin{equation}\label{d12}
{1 \ \ \,0\choose 0 \ \ \,1}D^{1/2}_{s,c}(R^W_{L_{\bf\Delta}})=
S^{-1}({\bf p}_{\bar s,c})S({\bf\Delta})S({\bf p}),
\end{equation}
where
$$
S({\bf p})=\sqrt{\frac{\epsilon(p)+m}{2m}}\left(1+\frac{\bm{\alpha}{\bf p}}
{\epsilon(p)+m}\right)
$$
is the usual Lorentz transformation matrix of the Dirac spinor.

\section{Form factors of weak $B_s$ decays to $D_s$ mesons}
\label{sec:ffsdgs}
For considering weak $B_s$ decays to ground state $D_s$
mesons we employ the heavy quark expansion which significantly
simplifies calculations. Therefore it is convenient to introduce the
heavy quark effective theory (HQET)
parametrization for the weak decay matrix elements \cite{iw,n}:
\begin{eqnarray}\label{ff}
 {\langle D_s(v')| \bar c\gamma^\mu b |B_s(v)\rangle  
\over\sqrt{M_{D_s}M_{B_s}}}
  &=& h_+ (v+v')^\mu + h_- (v-v')^\mu , \\ \cr\label{ff1}
  \langle D_s(v')| \bar c\gamma^\mu b \gamma_5 |B_s(v)\rangle 
  &=& 0, \\ \cr\label{ff2}
  {\langle D^*_s(v',\epsilon)| \bar c\gamma^\mu b |B_s(v)\rangle  
\over\sqrt{M_{D^*_s}M_{B_s}}}
  &=& i h_V \varepsilon^{\mu\alpha\beta\gamma} 
  \epsilon^*_\alpha v'_\beta v_\gamma ,\\ \cr\label{ff3}
  {\langle D^*_s(v',\epsilon)| \bar c\gamma^\mu\gamma_5 b |B_s(v)\rangle  
\over\sqrt{M_{D^*_s}M_{B_s}}}
  &=& h_{A_1}(w+1) \epsilon^{* \mu} 
   -(h_{A_2} v^\mu + h_{A_3} v'^\mu) (\epsilon^*\cdot v) ,
   \end{eqnarray}
where $v~(v')$ is the four-velocity of the $B_s~(D^{(*)}_s)$ meson,
$\epsilon^\mu$ is the polarization vector of the final vector meson,
and the form
factors   $h_i$ are dimensionless functions of the product of
four-velocities $$w=v\cdot
v'=\frac{M_{B_s}^2+M_{D_s^{(*)}}^2-q^2}{2M_{B_s}M_{D_s^{(*)}}},$$
and $q=p_{B_s}-p_{D_s^{(*)}}$ is the momentum transfer from the parent to
daughter meson,  $M_{B_s}$ is the $B_s$ meson mass, $M_{D_s^{(*)}}$ is the final
$D_s^{(*)}$ meson mass  
and $\epsilon_\mu$ is the polarization vector of the
final vector $D^*_s$ meson. 

In HQET these form factors up to $1/m_Q$ order are expressed through one
leading  Isgur-Wise function $\xi$, four sudleading functions
$\xi_3$, $\chi_{1,2,3}$ and one mass parameter $\bar\Lambda$ \cite{n}. These
relations are given in Appendix~\ref{sec:ffg}.

To calculate the weak decay matrix element in the quasipotential
approach, we substitute the vertex functions (\ref{gamma1}) and
(\ref{gamma2}) in Eq.~(\ref{mxet}) and take into account
the wave function transformations (\ref{wig}). The contribution of the
leading order vertex function $\Gamma_\mu^{(1)}({\bf p},{\bf q})$ can
be easily simplified by carrying out one of the integrations using the
$\delta$-function.  Then we employ the
heavy quark expansion, which permits us to take one of the integrals in
the contribution of the vertex function $\Gamma_\mu^{(2)}({\bf p},{\bf
  q})$ to the weak current matrix element. As a result we express all
matrix elements through the usual overlap integrals of the meson wave
functions. We carry out the heavy
quark expansion up to the second order and compare  the obtained
expressions with  model independent HQET relations
(\ref{cffgs})-(\ref{cffgsl}).  

All leading order relations are exactly satisfied.
In this limit of an infinitely heavy
quark, all form factors are expressed through the single universal Isgur-Wise function $\xi(w)$
\cite{iw} 
\begin{eqnarray}
&&h_+(w)=h_{A_1}(w)=h_{A_3}(w)=h_{V}(w)=\xi(w)\cr
&&h_{-}(w)=h_{A_2}(w)=0.
\end{eqnarray}
This function is given by the following overlap integral of
meson wave functions  \cite{fg}
\begin{equation} \label{iw}
  \xi(w)=\sqrt{\frac{2}{w+1}}\lim_{m_Q\to\infty} \int
  \frac{d^3p}{(2\pi)^3} \bar\Psi_{D_s}\!\!\left({\bf p}+
    2\epsilon_s(p)\sqrt{\frac{w-1}{w+1}} {\bf
      e_\Delta}\right)\Psi_{B_s}({\bf p}),
\end{equation} 
where $ {\bf e_\Delta}={\bf \Delta}/\sqrt{{\bf \Delta}^2}$ is the unit
vector in the direction of ${\bf \Delta}=M_{D_s}{\bf v}'-M_{B_s}{\bf
  v}$. In the infinitely
heavy quark mass limit the wave functions of initial $\Psi_{B_s}$ and
final $\Psi_{D_s}$ heavy mesons coincide. As a result the HQET
normalization condition \cite{n} $$\xi(1)=1$$ is exactly
reproduced.

 In order
to fulfill the HQET relations (\ref{cffgs})-(\ref{cffgsl}) at the first order of the heavy
quark $1/m_Q$ expansion it is necessary to set
$(1-\varepsilon)(1+\kappa)=0$, which leads to the vanishing long-range
chromomagnetic interaction.  This condition is satisfied by our choice
of the anomalous chromomagnetic quark moment $\kappa=-1$. To reproduce the HQET
relations at second order in  $1/m_Q$, one needs to set
$\varepsilon=-1$ \cite{fg}. This serves as an additional justification, based
on the heavy quark symmetry and heavy quark expansion in QCD, for the
choice of the characteristic parameters in our model.
    The  subleading Isgur-Wise functions are given by
\cite{fg,hlsem}  
\begin{eqnarray}
\label{fo}
\xi_3(w) &=& (\bar\Lambda -m_q) \left(1+
\frac{2}{3}\frac{w-1}{w+1}\right)\xi(w), \\ \label{fo1}
\chi_1(w)&=&\bar\Lambda\frac{w-1}{w+1}\xi(w), \\ \label{fo2}
\chi_2(w)&=&-\frac{1}{32}\frac{\bar\Lambda}{w+1}\xi(w), \\ \label{fo3}
\chi_3(w)&=&\frac{1}{16}\bar\Lambda\frac{w-1}{w+1}\xi(w),
\end{eqnarray}
where the HQET parameter $\bar \Lambda =M-m_Q$ is equal to the mean energy
of a light $s$ quark in a heavy meson $\langle \epsilon_s\rangle.$
The functions $\chi_1$ and $\chi_3$ explicitly satisfy normalization
conditions at the zero recoil point  \cite{luke} 
$$\chi_1(1)=\chi_3(1)=0,$$
arising from the vector current conservation.

Near the zero recoil
point of the final meson $w=1$ the Isgur-Wise functions have the
following expansions  
\begin{eqnarray}
  \label{eq:iwexp}
  \xi(w)&=&1-1.466(w-1)+1.844(w-1)^2+\cdots,\cr\cr
 \xi_3(w)/\bar \Lambda&=&0.359-0.408(w-1)+0.428(w-1)^2+\cdots,\cr\cr
 \chi_1(w)/\bar \Lambda&=&0.499(w-1)-0.982(w-1)^2+\cdots,\cr\cr
 \chi_2(w)/\bar
 \Lambda&=&-0.0156+0.0307(w-1)-0.0442(w-1)^2+\cdots,\cr\cr
 \chi_3(w)/\bar \Lambda&=&0.0312(w-1)-0.0614(w-1)^2+\cdots
\end{eqnarray}

\begin{figure}
\centering
  \includegraphics[height=5.2cm]{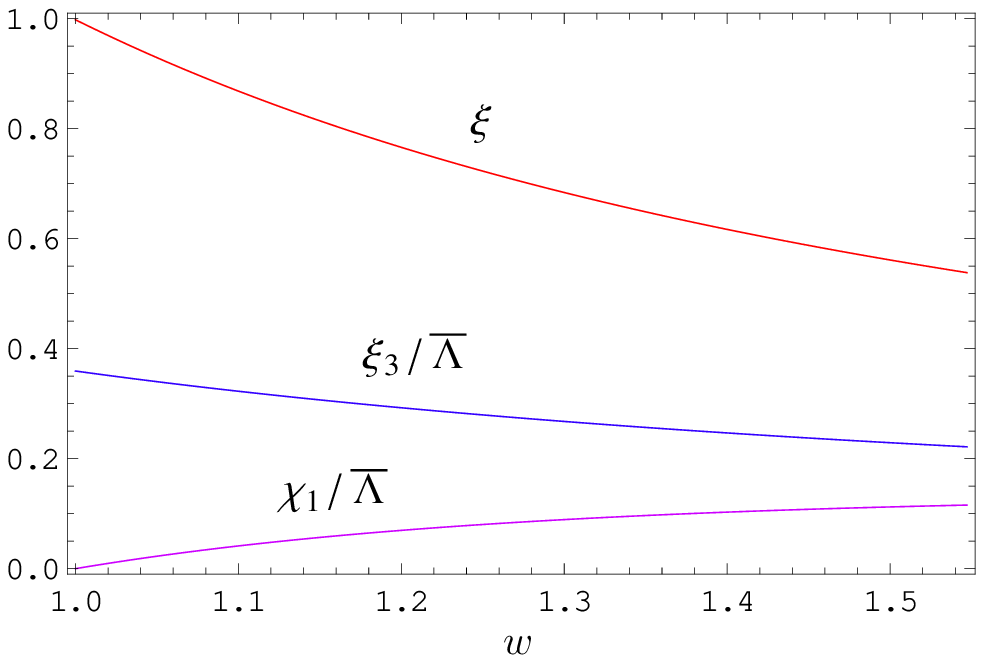}\ \
 \ \includegraphics[height=5.2cm]{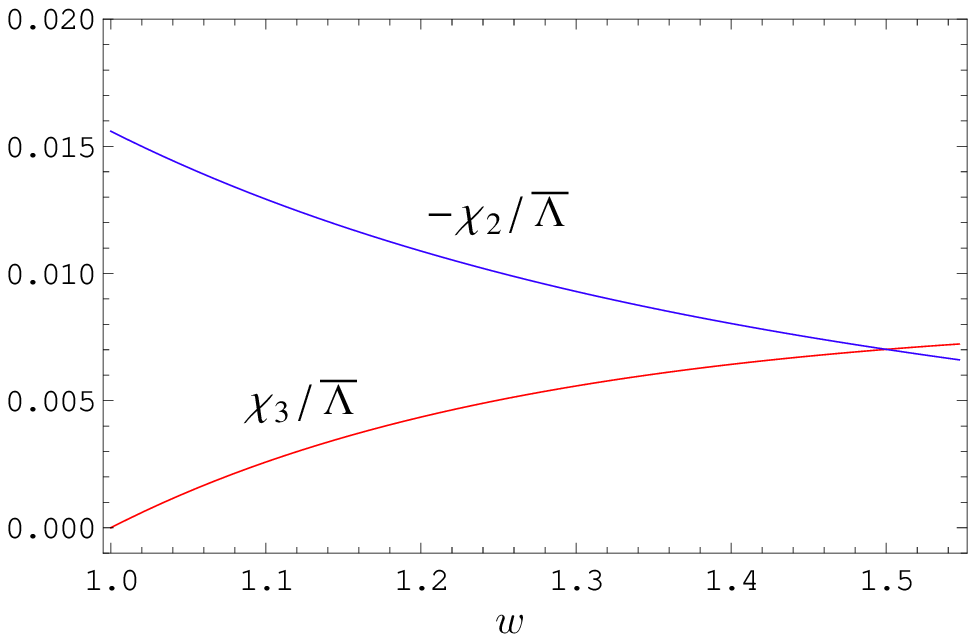}\\
\caption{Leading and subleading Isgur-Wise functions for $B_s\to D_s^{(*)}$ transitions.    } 
\label{fig:iw}
\end{figure}

The calculated leading and subleading Isgur-Wise functions for $B_s\to
D_s^{(*)}$ transitions are plotted in Fig.~\ref{fig:iw}. 
Using these Isgur-Wise functions we obtain the decay form factors
$h_i(w)$ with the account of the first order $1/m_Q$ corrections in the
whole kinematical range. To improve calculations we scale the results
by the values of the form factors at zero recoil $h_i(1)$ evaluated
with the inclusion of $1/m_Q^2$ corrections using formulas given in
Ref.~\cite{fg}. We find that the account of the  first order
corrections changes the form factors by about 17\%, while the
contribution of the second order corrections is less than
3\%. These values are in accord with the naive estimates of such
corrections $\bar \Lambda/(2m_c)\approx 0.15$ and $[\bar
\Lambda/(2m_c)]^2\approx 0.02$. 

The other popular parametrization for 
the matrix elements of weak current $J^W$ between meson states is
given by
\begin{equation}
  \label{eq:pff1}
  \langle D_s(p_{D_s})|\bar c \gamma^\mu b|B_s(p_{B_s})\rangle
  =f_+(q^2)\left[p_{B_s}^\mu+ p_{D_s}^\mu-
\frac{M_{B_s}^2-M_{D_s}^2}{q^2}\ q^\mu\right]+
  f_0(q^2)\frac{M_{B_s}^2-M_{D_s}^2}{q^2}\ q^\mu,
\end{equation}
\begin{equation}
\label{eq:pff2} 
 \langle D_s(p_{D_s})|\bar c \gamma^\mu\gamma_5 b|B_s(p_{B_s})\rangle
  =0,
\end{equation}
\begin{eqnarray}
  \label{eq:vff1}
  \langle {D^*_s}(p_{D^*_s})|\bar c \gamma^\mu b|B(p_{B_s})\rangle&=
  &\frac{2iV(q^2)}{M_{B_s}+M_{D^*_s}} \epsilon^{\mu\nu\rho\sigma}\epsilon^*_\nu
  p_{B_s\rho} p_{{D^*_s}\sigma},\\ \cr
\label{eq:vff2}
\langle {D^*_s}(p_{D^*_s})|\bar c \gamma^\mu\gamma_5 b|B_s(p_{B_s})\rangle&=&2M_{D^*_s}
A_0(q^2)\frac{\epsilon^*\cdot q}{q^2}\ q^\mu
 +(M_{B_s}+M_{D^*_s})A_1(q^2)\left(\epsilon^{*\mu}-\frac{\epsilon^*\cdot
    q}{q^2}\ q^\mu\right)\cr\cr
&&-A_2(q^2)\frac{\epsilon^*\cdot q}{M_{B_s}+M_{D^*_s}}\left[p_{B_s}^\mu+
  p_{D^*_s}^\mu-\frac{M_{B_s}^2-M_{D^*_s}^2}{q^2}\ q^\mu\right]. 
\end{eqnarray}
At the maximum recoil point ($q^2=0$) these form
factors satisfy the following conditions: 
\[f_+(0)=f_0(0),\]
\[A_0(0)=\frac{M_{B_s}+M_{D^*_s}}{2M_{D^*_s}}A_1(0)
-\frac{M_{B_s}-M_{D^*_s}}{2M_{D^*_s}}A_2(0).\]

The relations between two sets of weak decay form factors are given in Appendix~\ref{ffr}.

Substituting in these relations the Isgur-Wise functions of our
model (\ref{iw})-(\ref{fo3}) we find that the decay form factors can
be approximated with sufficient accuracy by the following expressions:

(a) $f_+(q^2),V(q^2),A_0(q^2)=F(q^2)$ 
\begin{equation}
  \label{fitfv}
  F(q^2)=\frac{F(0)}{\displaystyle\left(1-\frac{q^2}{ M^2}\right)
    \left(1-\sigma_1 
      \frac{q^2}{M_{B_c^*}^2}+ \sigma_2\frac{q^4}{M_{B_c^*}^4}\right)},
\end{equation}

(b) $f_0(q^2), A_1(q^2),A_2(q^2)=F(q^2)$
\begin{equation}
  \label{fita12}
  F(q^2)=\frac{F(0)}{\displaystyle \left(1-\sigma_1
      \frac{q^2}{M_{B_c^*}^2}+ \sigma_2\frac{q^4}{M_{B_c^*}^4}\right)},
\end{equation}
where $M=M_{B_c^*}=6.332$~GeV  for the form factors $f_+(q^2),V(q^2)$ and
$M=M_{B_c}=6.272$~GeV for the form factor $A_0(q^2)$; the values  $F(0)$ and $\sigma_{1,2}$ are given in 
Table~\ref{hff}. The values
of $\sigma_{1,2}$ are determined with a few tenths of percent
errors. The main uncertainties of the form factors originate from 
the account of $1/m_Q^2$ corrections at zero recoil only and from
the higher order $1/m_Q^3$ contributions and can be roughly
estimated in our approach to be about 2\%.~{\footnote{Other
uncertainties originating, e.g., from meson wave functions and model
parameters are significantly smaller. Indeed meson wave functions and
masses were obtained by the numerical solution of the quasipotential
equation with the completely relativistic spin-independent and
spin-dependent potentials treated nonperturbatively \cite{hlm}. The
model parameters were fixed in previous calculations which correctly
reproduce numerous experimental data. The integrated quantities such
as decay form factors and 
semileptonic decay rates are much less sensitive to the variation
of the model parameters than such quantities as hadron masses
which are measured with considerably higher accuracy. Thus
even the limited variation of these parameters, permitted by the
description of hadron masses, will give significantly smaller
contributions to the form factor and  decay rate uncertainties
compared to the ones mentioned above.} The $q^2$ dependence of
these form factors is shown in Fig.~\ref{fig:fplf0}.

\begin{table}
\caption{Form factors of weak $B_s\to D_s^{(*)}$ transitions 
  calculated in our model. Form factors $f_+(q^2)$, $V(q^2)$,
  $A_0(q^2)$ are fitted by Eq.~(\ref{fitfv}), and form factors $f_0(q^2)$,
  $A_1(q^2)$, $A_2(q^2)$ are fitted by Eq.~(\ref{fita12}).  }
\label{hff}
\begin{ruledtabular}
\begin{tabular}{ccccccc}
   &\multicolumn{2}{c}{{$B_s\to D_s$}}&\multicolumn{4}{c}{{\  $B_s\to D^*_s$\
     }}\\
\cline{2-3} \cline{4-7}
& $f_+$ & $f_0$& $V$ & $A_0$ &$A_1$&$A_2$ \\
\hline
$F(0)$          &0.74&0.74 &  0.95 & 0.67 & 0.70& 0.75\\
$F(q^2_{\rm max})$&1.15&0.88 &  1.50 & 1.06& 0.84& 1.04 \\
$\sigma_1$      &0.200&0.430& 0.372 &0.350&  0.463& 1.04\\
$\sigma_2$      &$-0.461$&$-0.464$&$-0.561$&$-0.600$&$-0.510$&$-0.070$\\
\end{tabular}
\end{ruledtabular}
\end{table}

\begin{figure}
\centering
  \includegraphics[width=8cm]{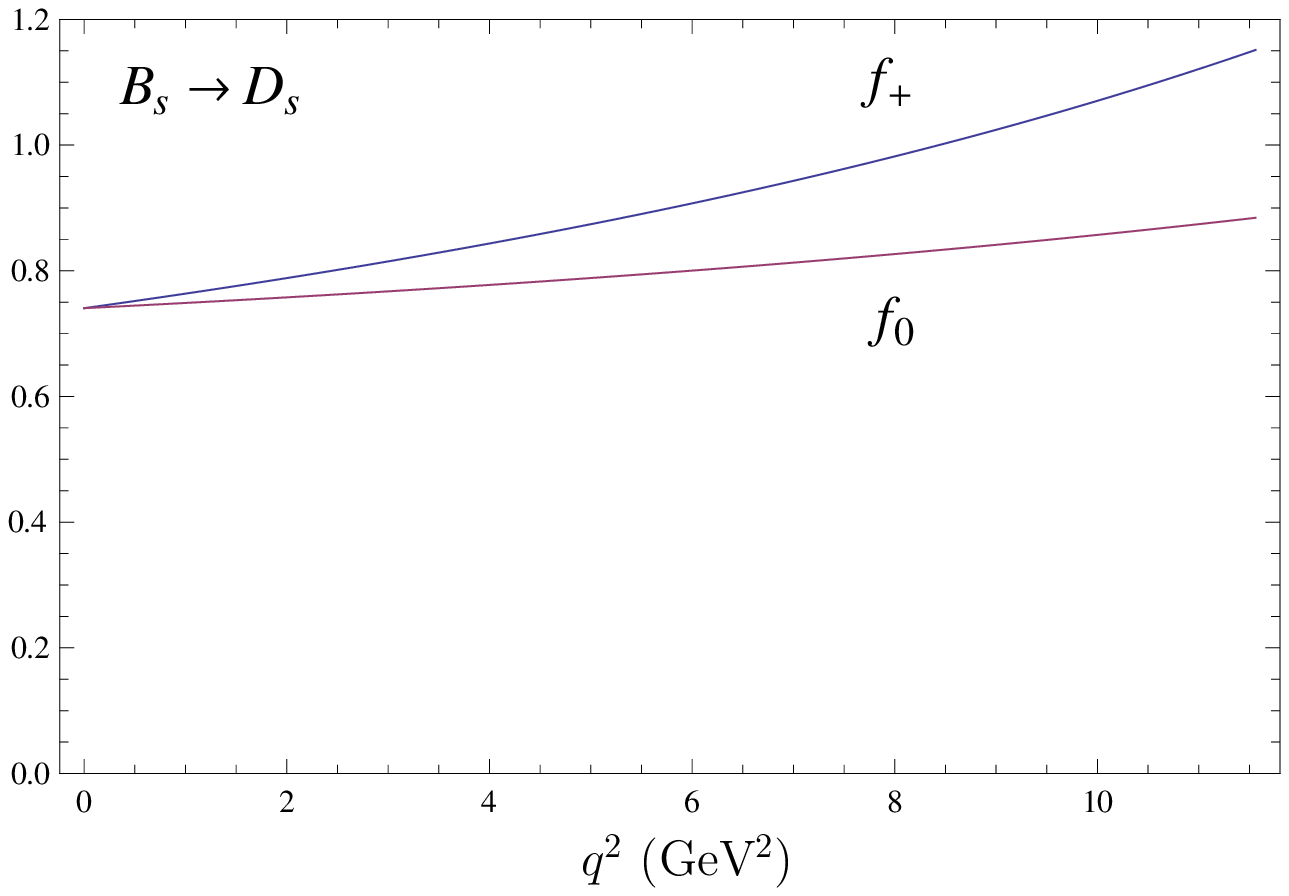}\ \
 \ \includegraphics[width=8cm]{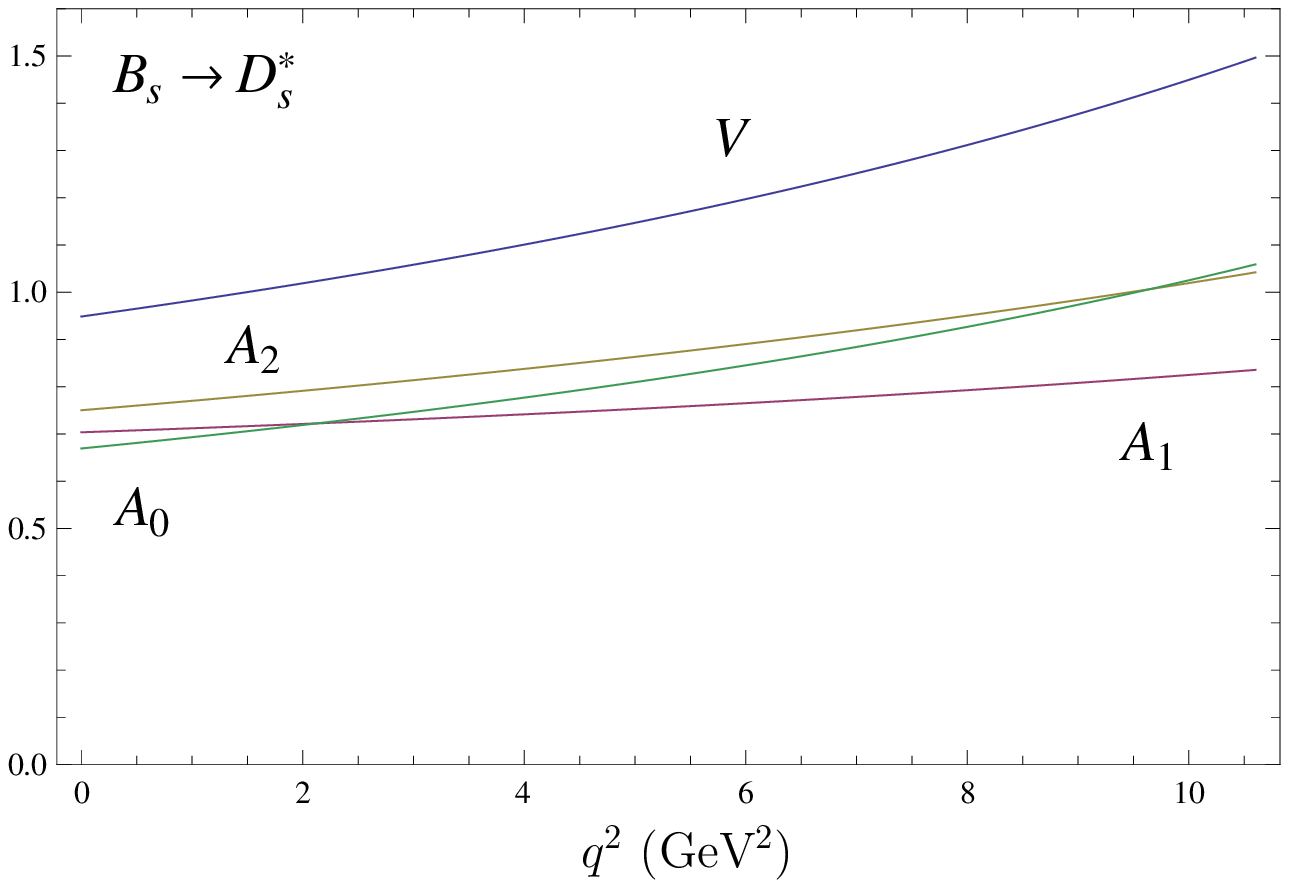}\\
\caption{Form factors of  weak $B_s\to D_s^{(*)}$ transitions.    } 
\label{fig:fplf0}
\end{figure}

\begin{table}
\caption{Comparison of theoretical predictions for the form factors of 
  semileptonic decays $B_s\to  D_s^{(*)} e\nu$  at maximum
  recoil point $q^2=0$.  }
\label{compbpiff}
\begin{ruledtabular}
\begin{tabular}{cccccc}
     & $f_+(0)$ & $V(0)$ & $A_0(0)$ &$A_1(0)$&$A_2(0)$ \\
\hline
this paper      &$0.74\pm0.02$  & $0.95\pm0.02$  & $0.67\pm0.01$  & $0.70\pm0.01$ & $0.75\pm0.02$\\
\cite{kp}&0.61  & 0.64  &       & 0.56 & 0.59\\
\cite{bcnp} &$0.7\pm0.1$ &$0.63\pm0.05$ &$0.52\pm0.06$ &$0.62\pm0.01$&$0.75\pm0.07$ \\
\cite{cfkw} &$0.57^{+0.02}_{-0.03}$ &$0.70^{+0.05}_{-0.04}$ &
&$0.65^{+0.01}_{-0.01}$&$0.67^{+0.01}_{-0.01}$  \\
\cite{llw} & $0.86^{+0.17}_{-0.15}$\\
\cite{lsw} & &$0.74^{+0.05}_{-0.05}$ &$0.63^{+0.04}_{-0.04}$ 
&$0.61^{+0.04}_{-0.04}$&$0.59^{+0.04}_{-0.04}$
\end{tabular}
\end{ruledtabular}
\end{table} 

In Table \ref{compbpiff} we confront our predictions for the form factors of 
semileptonic decays $B_s\to  D_s^{(*)} e\nu$  at maximum recoil point
$q^2=0$ with results of other approaches
\cite{kp,bcnp,cfkw,llw,lsw}. Different quark models are used in
Refs.~\cite{kp,cfkw,lsw}, while the QCD and light cone sum rules are
employed in Refs.~\cite{bcnp,llw}. We find that these significantly
different theoretical calculations lead to rather close values of the
decay form factors. One of the main advantages of our model is its
ability not only to obtain the decay form factors at the single
kinematical point, but also to determine its $q^2$ dependence in the
whole range without any additional assumptions or extrapolations. 

\section{Semileptonic $B_s$ decays to $D_s$ mesons}
\label{sdgs}

The differential decay rate for the semileptonic $B_s$ meson decay to $D_s^{(*)}$ mesons reads \cite{iks}
\begin{equation}
  \label{eq:dgamma}
  \frac{d\Gamma(B_s\to D_s^{(*)}l\bar\nu)}{dq^2}=\frac{G_F^2}{(2\pi)^3}
  |V_{cb}|^2
  \frac{\lambda^{1/2}(q^2-m_l^2)^2}{24M_{B_s}^3q^2} 
  \Biggl[H H^{\dag}\left(1+\frac{m_l^2}{2q^2}\right)  +\frac{3m_l^2}{2q^2} H_tH^{\dag}_t\Biggr],
\end{equation}
where $G_F$ is the Fermi constant, $V_{cb}$ is the CKM matrix element, $\lambda\equiv
\lambda(M_{B_s}^2,M_{D_s^{(*)}}^2,q^2)=M_{B_s}^4+M_{D_s^{(*)}}^4+q^4-2(M_{B_s}^2M_{D_s^{(*)}}^2+M_{D_s^{(*)}}^2q^2+M_{B_s}^2q^2)$,
$m_l$ is the lepton mass and
\begin{equation}
  \label{eq:hh}
  H H^{\dag}\equiv H_+H^{\dag}_++H_-H^{\dag}_-+H_0H^{\dag}_0.
\end{equation}
The
helicity components $H_\pm$, $H_0$ and $H_t$ of the hadronic tensor are expressed through the
invariant form factors. They are given in Appendix~\ref{sec:hc}.

\begin{figure}
  \centering
 \includegraphics[width=8cm]{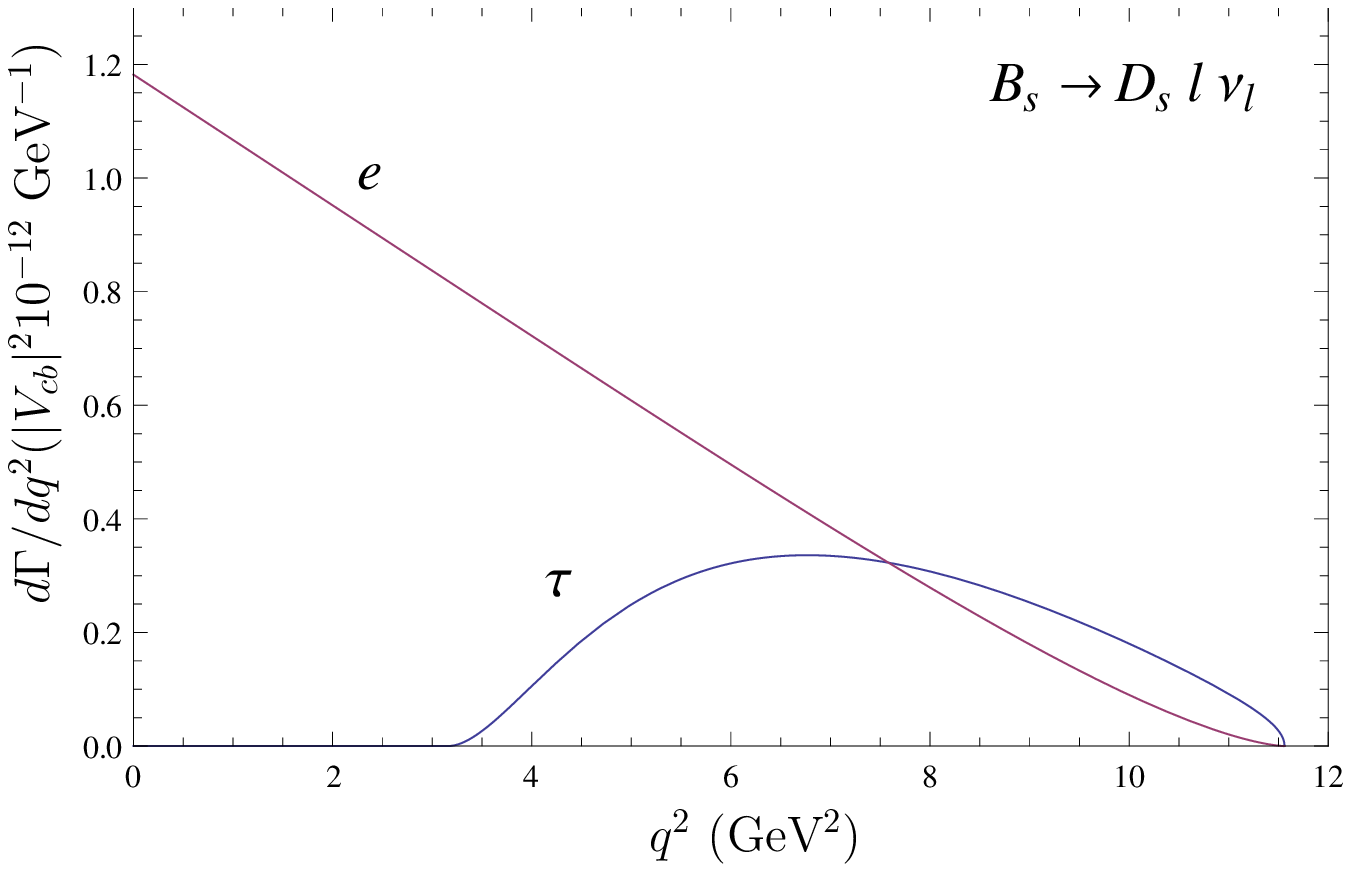}\ \
 \  \includegraphics[width=8cm]{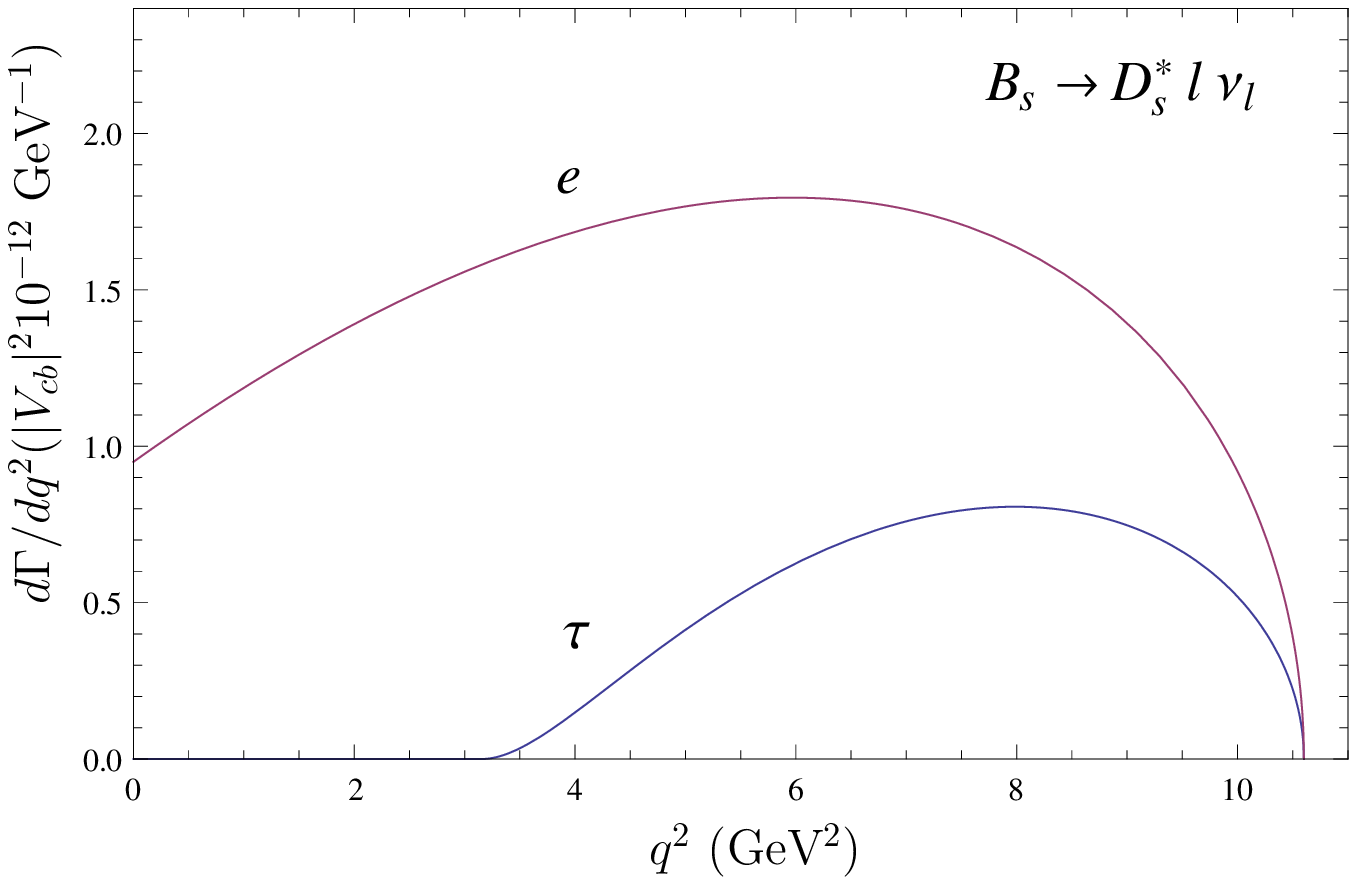}

  \caption{Predictions for the differential decay rates  of the $B_s\to D_s^{(*)}l\nu$ 
    semileptonic decays. }
  \label{fig:brbpr}
\end{figure}

Now we substitute the weak decay form factors calculated in the
previous section into the above expressions for decay
rates. The resulting differential decay rates for the $B_s$ decays to
the $D_s^{(*)}$ mesons
are plotted in Fig.~\ref{fig:brbpr}.  The corresponding total decay
rates are obtained  by integrating
the differential decay rates over $q^2$. For
calculations we use  the CKM matrix element
$|V_{cb}|=(3.9\pm0.15)\times 10^{-2}$, which was obtained from the
comparison of our theoretical predictions \cite{hlsem,fgtau} for the products
$F_{D^{(*)}}(w)|V_{cb}|$ and for the $B\to D^{(*)}l\nu_l$ decay branching
fractions with updated experimental data.~\footnote{This value of $|V_{cb}|$  is in
  accord with its recent evaluation by the Heavy Flavor
  Averaging Group \cite{hfag}.} It is necessary to point out that the kinematical
range  accessible in these semileptonic decays is rather
broad. Therefore the knowledge of the $q^2$ dependence of the form
factors is very important for reducing theoretical uncertainties of the decay
rates. Our results for the semileptonic $B_s\to D_s^{(*)} l\nu$ decay
rates are given in Table~\ref{comphlff} in comparison 
with previous calculations. The authors of Ref.\cite{bcnp} use the QCD
sum rules, while the light cone sum rules approach is adopted in
Ref.~\cite{llw}. Different types of constituent quark models are
employed in Refs.~\cite{lsw,cfkw,zll} and the three point QCD sum
rules are used in Ref.~\cite{ab}. We see that our predictions are
consistent with results of quark model calculations in
Refs.~\cite{lsw,cfkw}. They are approximately two times larger than
the QCD sum rules and light cone sum rules results of
Refs.~\cite{bcnp,llw}, but slightly lower than the values of
Refs.~\cite{zll,ab}. 

We find that the total branching fraction of the semileptonic decays of
$B_s$ mesons to the ground state $D_s^{(*)}$ is equal to
$Br(B_s\to D_s^{(*)} e\nu)=(7.4\pm 0.7)\%$ and $Br(B_s\to D_s^{(*)} \tau\nu)=(1.92\pm 0.15)\%$. The errors in our estimates originate from the
uncertainties in the determination of the CKM matrix element
$|V_{cb}|$, which are dominant, and from the theoretical uncertainties
in the determination of decay form factors. The latter uncertainties
are considerably smaller than the former ones and are
mostly related with the estimates of the higher order terms in the heavy
quark expansion.

\begin{table}
\caption{Comparison of theoretical predictions for the branching fractions of semileptonic
  decays $B_s\to D_s^{(*)} l\nu$ (in \%).  }
\label{comphlff}
\begin{ruledtabular}
\begin{tabular}{cccccccc}
Decay& this paper & \cite{bcnp}&\cite{cfkw}&\cite{llw}  &\cite{lsw} &\cite{zll}
 & \cite{ab}\\
\hline
$B_s\to D_se\nu$& $2.1\pm0.2$ & $1.35\pm0.21$ & 1.4-1.7&$1.0^{+0.4}_{-0.3}$&  &
2.73-3.00& 2.8-3.8\\ 
$B_s\to D_s\tau\nu$& $0.62\pm0.05$ & & 0.47-0.55 &$0.33^{+0.14}_{-0.11}$&  &\\
$B_s\to D_s^*e\nu$& $5.3\pm0.5$ & $2.5\pm0.1$& 5.1-5.8 && $5.2\pm0.6$  &
7.49-7.66& 1.89-6.61\\
$B_s\to D_s^*\tau\nu$& $1.3\pm0.1$ &  & 1.2-1.3&&$1.3^{+0.2}_{-0.1}$ &\\
\end{tabular}
\end{ruledtabular}
\end{table}

\section{Form factors of weak $B_s$ decays to radially excited $D_s(2S)$ mesons}
\label{sec:ffrexc}

The decay form factors (\ref{ff})-(\ref{ff3}) up to
$1/m_Q$ order in HQET for
$B_s$ decays to radially excited $D_s[(n+1)S]$ mesons are expressed through one
leading $\xi^{(n)}$ and five subleading $\tilde\xi_3$,
$\tilde\chi_{1,2,3,b}$ Isgur-Wise functions and two mass parameters
$\bar\Lambda$ and $\bar\Lambda^{(n)}$ \cite{radexc}. They are presented in Appendix~\ref{sec:ffre}.

In our model all HQET relations (\ref{cff})-(\ref{cff8}) are satisfied
and we get the following
expressions for the leading and subleading Isgur-Wise functions  \cite{radexc}:
\begin{eqnarray}
\label{xi}  
\xi^{(1)}(w)&=&\left(\frac{2}{w+1}\right)^{1/2}\int\frac{d^3 p}
{(2\pi)^3}\bar\psi^{(0)}_{D_s(2S)}\!\!\left({\bf p}+2\epsilon_s(p)\sqrt{\frac{w-1}{w+1}} {\bf
      e_\Delta}\right)
\psi^{(0)}_{B_s}({\bf p}),\\ \cr\cr
\label{xi3}
\tilde\xi_3(w)&=&\left(\frac{\bar\Lambda^{(1)}+\bar\Lambda}2-m_s+
\frac16\frac{\bar\Lambda^{(1)}-\bar\Lambda}{w-1}\right)\left(1+
\frac23\frac{w-1}{w+1}\right)\xi^{(1)}(w),\\ \cr
\label{chi1}
\tilde\chi_1(w)&\cong &\frac1{20}\frac{w-1}{w+1}\frac{\bar\Lambda^{(1)}
-\bar\Lambda}{w-1}\xi^{(1)}(w)\cr \cr
&&+\frac{\bar\Lambda^{(1)}}2
\left(\frac{2}{w+1}\right)^{1/2}\int\frac{d^3 p}
{(2\pi)^3}\bar\psi^{(1)si}_{D_s(2S)}\!\!\left({\bf p}+
 2\epsilon_s(p)\sqrt{\frac{w-1}{w+1}} {\bf
      e_\Delta}\right)
\psi^{(0)}_{B_s}({\bf p}),\\ \cr
\label{chi2}
\tilde\chi_2(w)&\cong& -\frac1{12}\frac{1}{w+1}\frac{\bar\Lambda^{(1)}
-\bar\Lambda}{w-1}\xi^{(1)}(w),\\ \cr
\label{chi3}
\tilde\chi_3(w)&\cong& -\frac3{80}\frac{w-1}{w+1}\frac{\bar\Lambda^{(1)}
-\bar\Lambda}{w-1}\xi^{(1)}(w)\cr \cr
&&+\frac{\bar\Lambda^{(1)}}4
\left(\frac{2}{w+1}\right)^{1/2}\int\frac{d^3 p}
{(2\pi)^3}\bar\psi^{(1)sd}_{D_s(2S)}\!\!\left({\bf p}+2\epsilon_s(p)\sqrt{\frac{w-1}{w+1}} {\bf
      e_\Delta}\right)
\psi^{(0)}_{B_s}({\bf p}),\\ \cr
\label{chib}
\chi_b(w)\!\!&\cong&\!\! \bar\Lambda\left(\frac{2}{w+1}\right)^{1/2}\!\int\frac{d^3 p}
{(2\pi)^3}\bar\psi^{(0)}_{D_s(2S)}\!\!\left({\bf p}+2\epsilon_s(p)\sqrt{\frac{w-1}{w+1}} {\bf
      e_\Delta}\right)
\left[\psi^{(1)si}_{B_s}({\bf p})-3\psi^{(1)sd}_{B_s}({\bf
    p})\right],\ \ \ \ \ \
\end{eqnarray}
where ${\bf \Delta}^2=M_{D_s^{(*)}(2S)}^2(w^2-1)$.
Here we used the expansion for the $S$-wave meson wave function  \cite{radexc}
$$\psi_M=\psi_M^{(0)}+\bar\Lambda_M\varepsilon_Q\left(\psi_M^{(1)si}
+d_M\psi_M^{(1)sd}\right)+O(1/m_Q^2),$$
where $\psi_M^{(0)}$ is the wave function in the limit $m_Q\to\infty$,
$\psi_M^{(1)si}$ and $\psi_M^{(1)sd}$ are the spin-independent and 
spin-dependent first order $1/m_Q$ corrections, $d_P=-3$ for pseudoscalar and
$d_V=1$ for vector mesons.
The symbol $\cong$ in the expressions (\ref{chi1})--(\ref{chib}) for the
subleading functions $\tilde\chi_i(w)$ implies that corrections
suppressed by an additional power of the ratio $(w-1)/(w+1)$, which is equal 
to zero  at $w=1$ and less than $1/6$ at $w_{\rm max}$, were neglected. 
Since the main contribution to the decay rate comes from the values of 
form factors  close to $w=1$, these corrections turn out to be unimportant. 
 
It is clear from the expression (\ref{xi}) that the leading order contribution
vanishes at the point of zero recoil (${\bf \Delta}=0, w=1$) of the 
final $D_s^{(*)}(2S)$ meson, 
since the radial parts of the wave functions $\Psi_{D_s(2S)}$ and $\Psi_{B_s}$
are orthogonal in the infinitely heavy quark limit.

Near the zero recoil
point of the final meson $w=1$ the Isgur-Wise functions have the
following expansions  
\begin{eqnarray}
  \label{eq:iwexpex}
  \xi^{(1)}(w)&=&2.455(w-1)-9.545(w-1)^2+\cdots,\cr\cr
 \tilde\xi_3(w)/\tilde \Lambda&=&0.221+1.094(w-1)-5.294(w-1)^2+\cdots,\cr\cr
 \tilde\chi_1(w)/\tilde \Lambda&=&0.182-0.143(w-1)-1.055(w-1)^2+\cdots,\cr\cr
 \tilde\chi_2(w)/\tilde
 \Lambda&=&-0.0552+0.242(w-1)-0.547(w-1)^2+\cdots,\cr\cr
 \tilde\chi_3(w)/\tilde
 \Lambda&=&-0.00133-0.0334(w-1)+0.150(w-1)^2+\cdots,\cr\cr
 \tilde\chi_b(w)/\tilde \Lambda&=&-0.169+1.114(w-1)-4.060(w-1)^2+\cdots,
\end{eqnarray}
where $\tilde \Lambda=(\bar\Lambda^{(1)}+\bar\Lambda)/2$.

The leading $\xi^{(1)}$ and subleading $\tilde\xi_3$,
$\tilde\chi_{1,2,3,b}$ Isgur-Wise functions for $B_s\to
D_s^{(*)}(2S)$ transitions are shown in Fig.~\ref{fig:iwe}. We use
relations (\ref{eq:ffiwpl})-(\ref{eq:ffiwa0}) to express form factors
$f_{+,0}(q^2)$, $V(q^2)$ and $A_{0,1,2}$ through the calculated
Isgur-Wise functions. The obtained form factors are plotted in
Fig.~\ref{fig:fpl2s}. Their values at zero and maximum $q^2$ are
given in Table~\ref{hlff}. The main theoretical uncertainties of the
decay from factors, as for the decays to the ground state mesons,
originate from the higher order $1/m_Q$ contributions and are less than 4\%.  Comparing plots in Figs.~\ref{fig:fplf0}
and \ref{fig:fpl2s} we see that form factors for the decays to ground
and radially excited states have significantly different behaviour
in $q^2$. The former ones grow with $q^2$, while the latter ones
decrease. This is the consequence of the different structure of nodes
of the wave functions of these states.    

\begin{figure}
\centering
  \includegraphics[width=8cm]{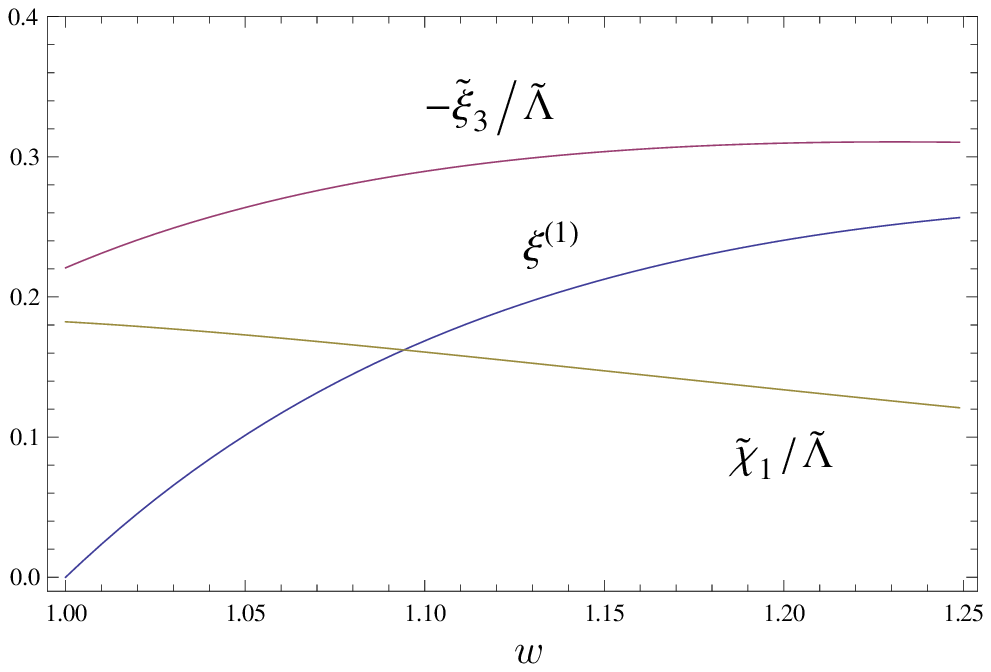}\ \
 \ \includegraphics[width=8cm]{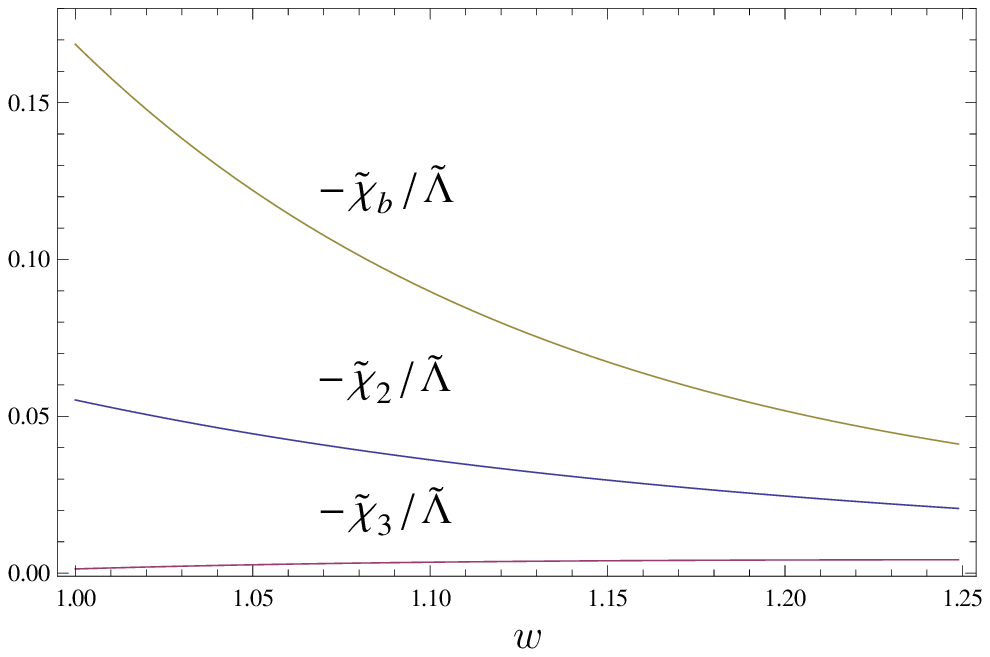}\\
\caption{Leading and subleading Isgur-Wise functions for $B_s\to
  D_s^{(*)}(2S)$ transitions ($\tilde \Lambda=(\bar\Lambda^{(1)}+\bar\Lambda)/2$).    } 
\label{fig:iwe}
\end{figure}

\begin{table}
\caption{Form factors of weak $B_s\to D_s^{(*)}(2S)$ transitions 
  calculated in our model.   }
\label{hlff}
\begin{ruledtabular}
\begin{tabular}{ccccccc}
   &\multicolumn{2}{c}{{$B_s\to D_s(2S)$}}&\multicolumn{4}{c}{{\  $B_s\to D^*_s(2S)$\
     }}\\
\cline{2-3} \cline{4-7}
& $f_+$ & $f_0$& $V$ & $A_0$ &$A_1$&$A_2$ \\
\hline
$F(0)$          &0.41&0.41 &  0.46 & 0.45 & 0.32& 0.058\\
$F(q^2_{\rm max})$&0.28&0.075 &  0.28 & 0.29& 0.072& $-0.29$ \\
\end{tabular}
\end{ruledtabular}
\end{table}

\begin{figure}
\centering
  \includegraphics[width=8cm]{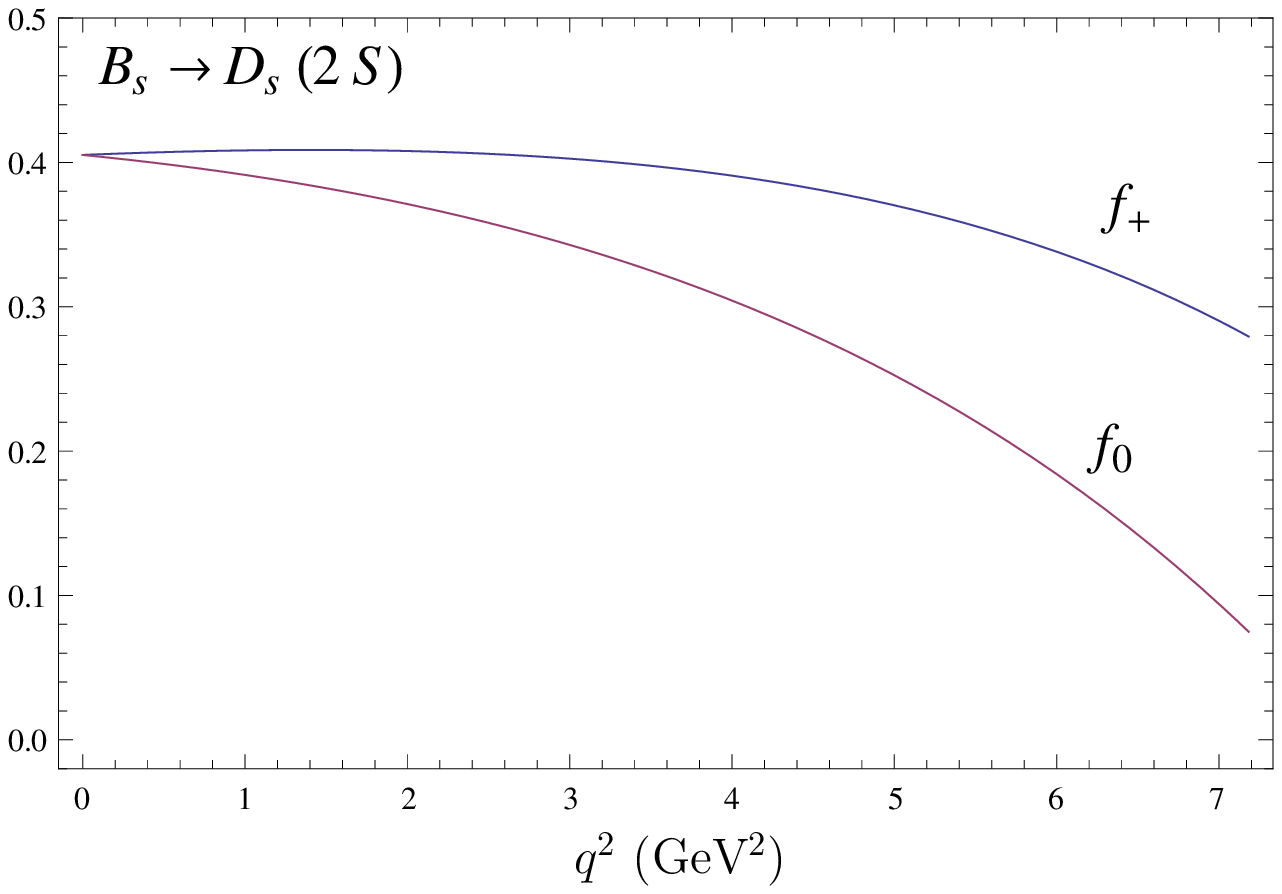}\ \
 \ \includegraphics[width=8.1cm]{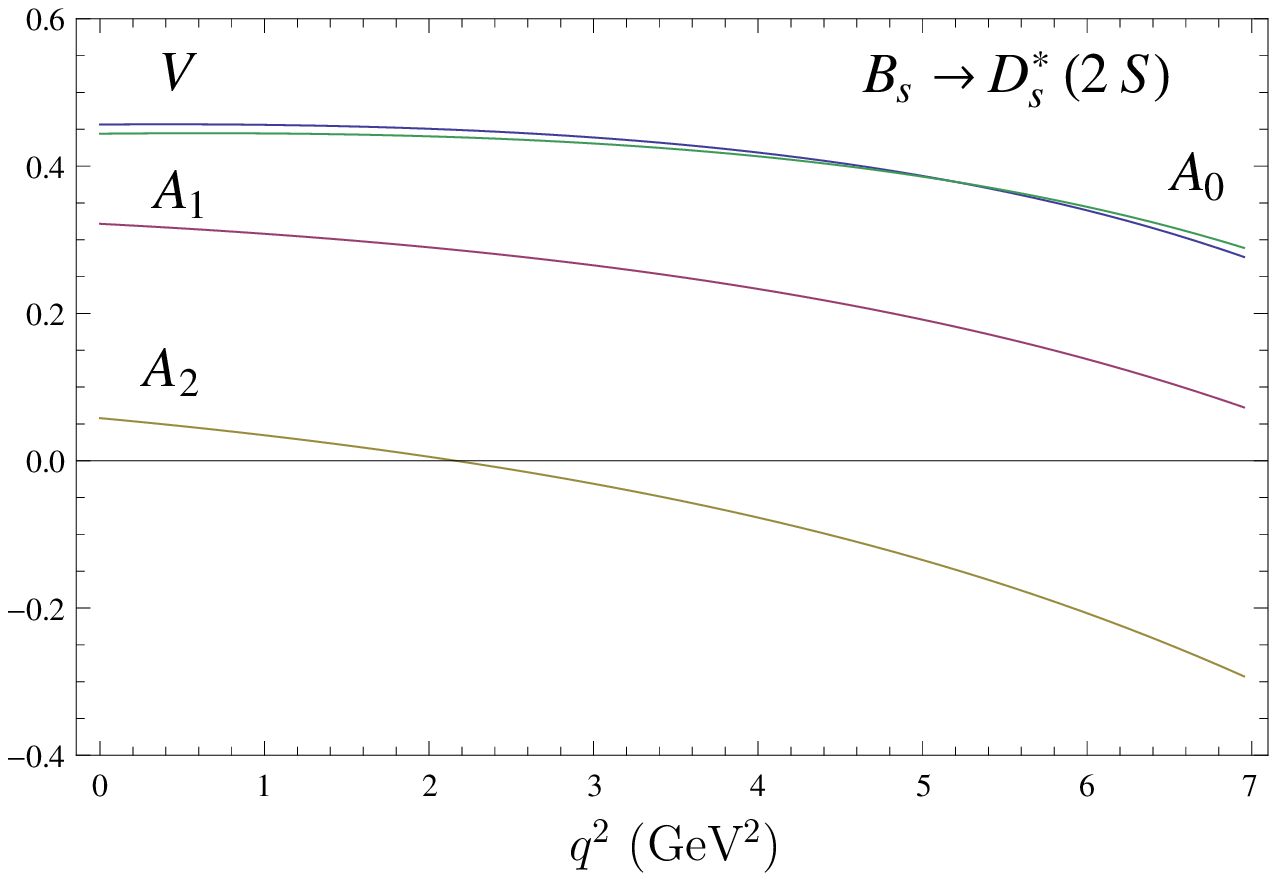}\\
\caption{Form factors of the weak $B_s\to D_s^{(*)}(2S)$ transitions.    } 
\label{fig:fpl2s}
\end{figure}

\section{Semileptonic $B_s$ decays to radially excited $D_s(2S)$
  mesons}
\label{sdre}

For the calculation of the semileptonic $B_s$ decays to radially
excited $D_s(2S)$ mesons we use the expression for the differential decay
rates (\ref{eq:dgamma}) with the helicity components of the hadronic
tensor given  by Eqs.~(\ref{eq:hsa})-(\ref{eq:hta}) and decay form
factors calculated in the previous section. The predictions
for the corresponding branching fractions are given in
Table~\ref{compreff}. We find that semileptonic $B_s$ decays to the
pseudoscalar $D_s(2S)$ and vector $D_s^*(2S)$ mesons have close values. The
total contribution of these decays is obtained to be $Br(B_s\to
D_s^{(*)}(2S)e\nu)=(0.65\pm0.06)\%$ and $Br(B_s\to
D_s^{(*)}(2S)\tau\nu)=(0.026\pm0.003)\%$.  

\begin{table}
\caption{Predictions for the branching fractions of semileptonic
  decays $B_s\to D_s^{(*)}(2S) l\nu$ (in \%).  }
\label{compreff}
\begin{ruledtabular}
\begin{tabular}{cc}
Decay& Br \\
\hline
$B_s\to D_s(2S)e\nu$& $0.27\pm0.03$ \\ 
$B_s\to D_s(2S)\tau\nu$& $0.011\pm0.001$\\
$B_s\to D_s^*(2S)e\nu$& $0.38\pm0.04$\\
$B_s\to D_s^*(2S)\tau\nu$& $0.015\pm0.002$ \\
\end{tabular}
\end{ruledtabular}
\end{table} 

\begin{figure}
  \centering
 \includegraphics[width=8cm]{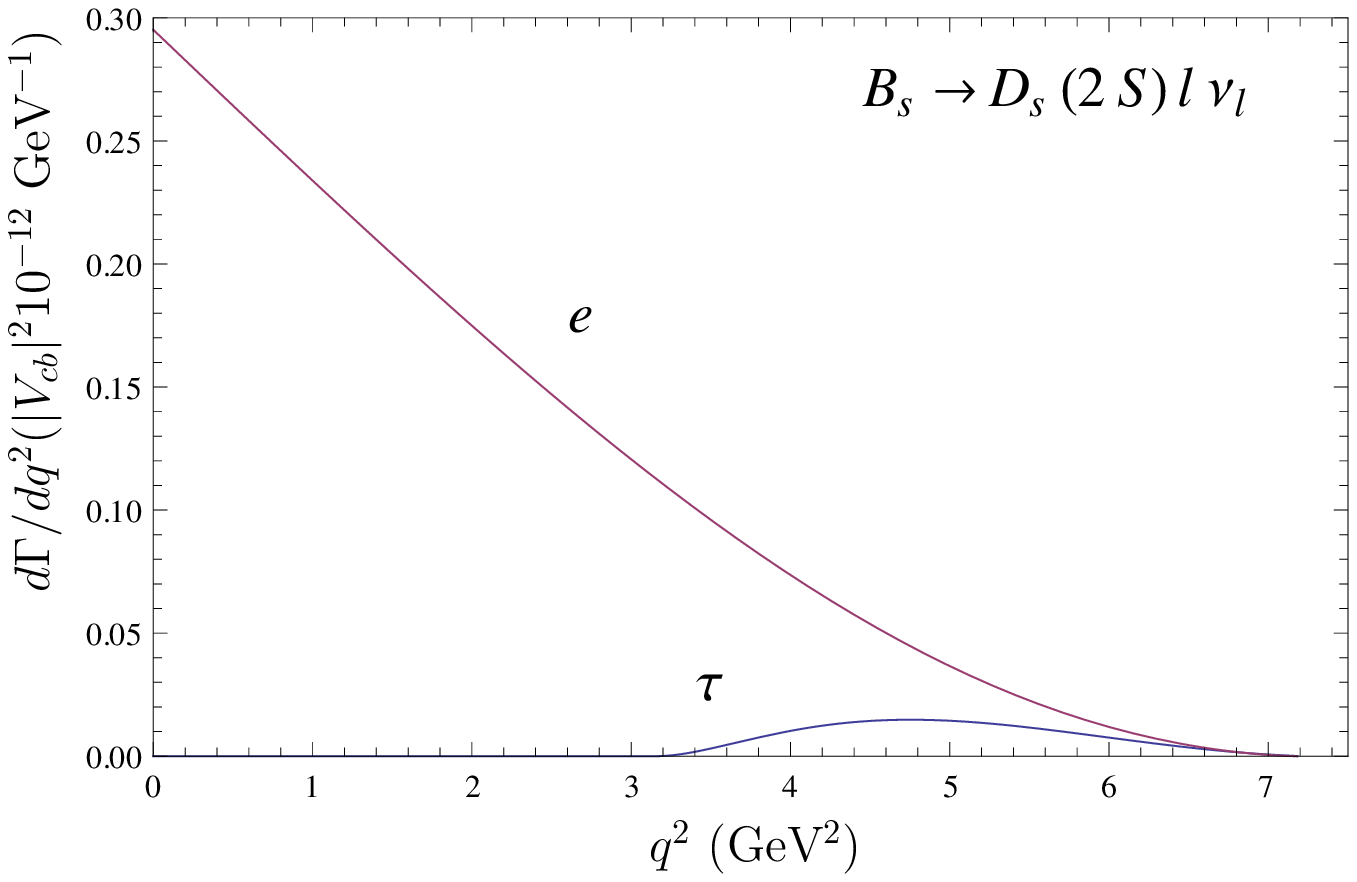}\ \
 \  \includegraphics[width=8cm]{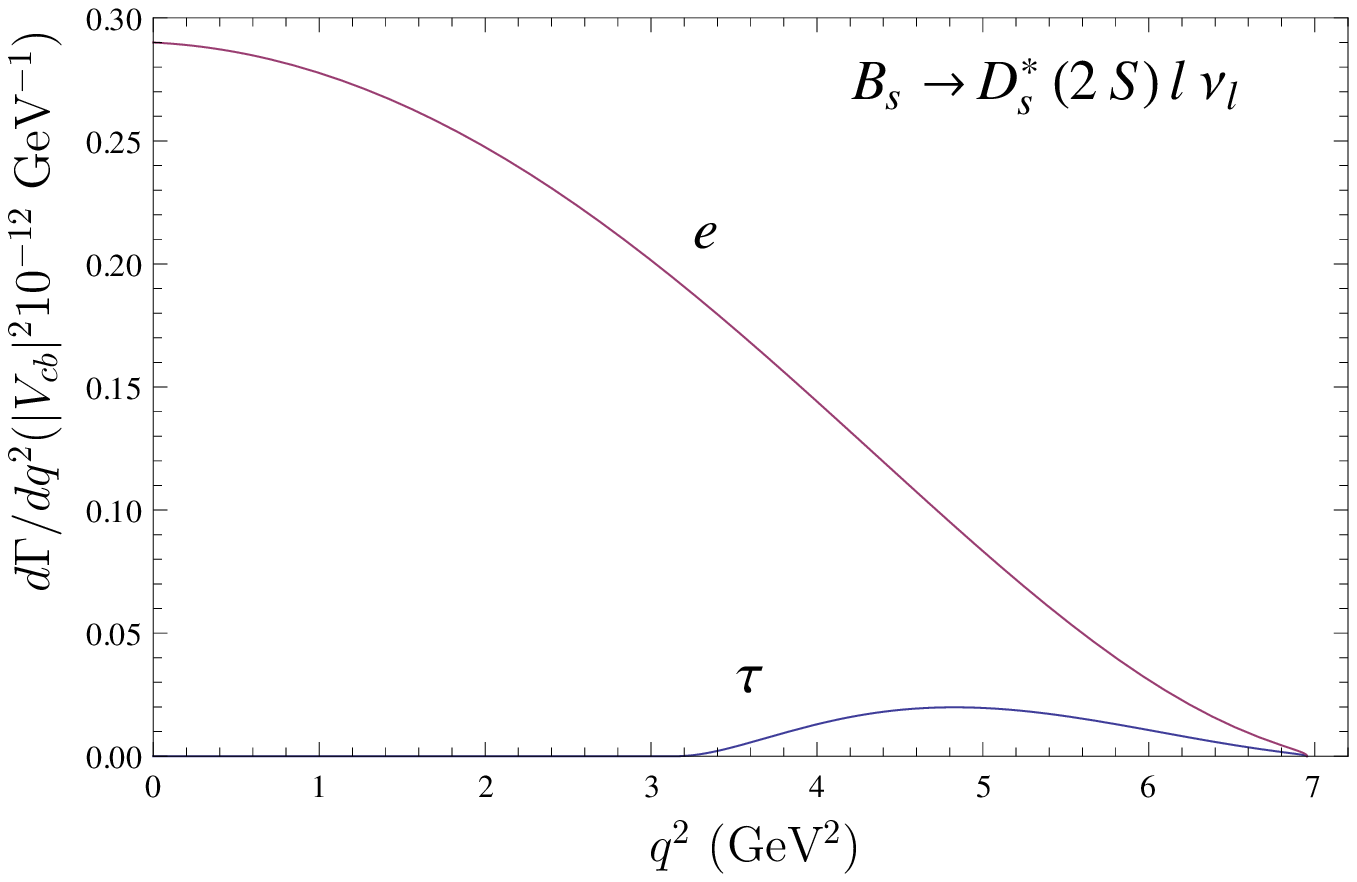}
\caption{Predictions for the differential decay rates  of the $B_s\to D_s^{(*)}(2S)l\nu$ 
    semileptonic decays. }
  \label{fig:brbprexc}

\end{figure}
The differential decay rates  of the $B_s\to D_s^{(*)}(2S)l\nu$  
semileptonic decays are plotted in Fig~\ref{fig:brbprexc}.

\section{Form factors of weak $B_s$ decays to orbitally excited $D_{sJ}^{(*)}$ mesons}
\label{sec:fforbexc}

The matrix elements of the weak current $J^W_\mu=\bar
b\gamma_\mu(1-\gamma_5)c$ for $B_s$ decays to orbitally
excited scalar $D_{s0}^*$ mesons can be parametrized by two invariant
form factors
\begin{eqnarray}
  \label{eq:sff1}
\langle D_{s0}^*(p_{D_{s0}})|\bar c \gamma^\mu b|B_s(p_{B_s})\rangle
  &=&0,\\ \cr
  \langle D_{s0}^*(p_{D_{s0}})|\bar c \gamma^\mu\gamma_5 b|B(p_{B_s})\rangle
  &=&r_+(q^2)\left(p_{B_s}^\mu+ p_{D_{s0}}^\mu\right)+
  r_-(q^2)\left(p_{B_s}^\mu- p_{D_{s0}}^\mu\right),
\end{eqnarray}
where $q=p_{B_s}-p_{D_{s0}}$, $M_{D_{s0}}$ is the scalar meson mass.

The matrix elements of the weak current for $B_s$ decays to the  axial
vector $D_{s1}$ meson 
can be expressed in terms of four invariant form factors
\begin{eqnarray}
  \label{eq:avff1}
  \langle D_{s1}(p_{D_{s1}})|\bar c \gamma^\mu b|B(p_{B_s})\rangle\!\!&=&\!\!
  (M_{B_s}+M_{D_{s1}})h_{V_1}(q^2)\epsilon^{*\mu}
  +[h_{V_2}(q^2)p_{B_s}^\mu+h_{V_3}(q^2)p_{D_{s1}}^\mu]\frac{\epsilon^*\cdot q}{M_{B_s}} ,\qquad\\\cr
\label{eq:avff2}
\langle D_{s1}(p_{D_{s1}})|\bar c \gamma^\mu\gamma_5 b|B(p_{B_s})\rangle\!\!&=&\!\!
\frac{2ih_A(q^2)}{M_{B_s}+M_{D_{s1}}} \epsilon^{\mu\nu\rho\sigma}\epsilon^*_\nu
  p_{B_s\rho} p_{{D_{s1}}\sigma},  
\end{eqnarray}
where $M_{D_{s1}}$ and $\epsilon^\mu$ are the mass and polarization vector of 
the axial vector meson. The matrix elements of the weak current for
$B_s$ decays to the axial vector $D_{s1}'$ meson are
obtained from Eqs.~(\ref{eq:avff1}) by the replacement of the set of
form factors $h_i(q^2)$ by $g_i(q^2)$ ($i=V_1,V_2,V_3,A$).  

The matrix elements of the weak current for $B_s$ decays to the tensor
$D_{s2}^*$ meson 
can be decomposed in four Lorentz-invariant structures
\begin{eqnarray}
  \label{eq:tff1}
  \langle D_{s2}^*(p_{D_{s2}})|\bar c \gamma^\mu b|B(p_{B_s})\rangle&=&
\frac{2it_V(q^2)}{M_{B_s}+M_{D_{s2}}} \epsilon^{\mu\nu\rho\sigma}\epsilon^*_{\nu\alpha}
\frac{p_{B_s}^\alpha}{M_{B_s}}  p_{B_s\rho} p_{{D_{s2}}\sigma},\\\cr
\label{eq:tff2}
\langle D_{s2}^*(p_{D_{s2}})|\bar q \gamma^\mu\gamma_5 b|B(p_{B_s})\rangle&=&
(M_{B_s}+M_{D_{s2}})t_{A_1}(q^2)\epsilon^{*\mu\alpha}\frac{p_{B_s\alpha}}{M_{B_s}}\cr\cr
&&  +[t_{A_2}(q^2)p_{B_s}^\mu+t_{A_3}(q^2)p_{D_{s2}}^\mu]\epsilon^*_{\alpha\beta}
\frac{p_{B_s}^\alpha p_{B_s}^\beta}{M_{B_s}^2} , 
\end{eqnarray}
where $M_{D_{s2}}$ and $\epsilon^{\mu\nu}$ are the mass and polarization tensor of 
the tensor meson.

To obtain  the form factors of $B_s\to D_{sJ}^{(*)}$
weak transitions we use the expression for the weak current matrix element
(\ref{mxet}). We calculate exactly the 
contribution of the leading vertex function $\Gamma^{(1)}({\bf p},{\bf q})$ 
(\ref{gamma1}) to the transition matrix element of the weak
current (\ref{mxet}) using the $\delta$-function.  For the evaluation of
the subleading contribution $\Gamma^{(2)}({\bf p},{\bf q})$ for the $B\to
D_{sJ}^{(*)}$ transitions, governed by the heavy-to-heavy
$b\to c$ transitions,  we use expansions in inverse powers of masses
of the heavy $b$- and $c$-quarks, contained in the initial $B_s$ meson and
final $D_{sJ}^{(*)}$ meson. Thus we can neglect the small
relative quark momentum $|{\bf p}|$ compared to the heavy quark mass $m_Q$
in the quark energy $\epsilon_Q(p+\Delta)\equiv\sqrt{m_Q^2+({\bf
    p}+{\bf\Delta})^2}$, replacing it by
$\epsilon_Q(\Delta)\equiv\sqrt{m_Q^2+{\bf\Delta}^2}$ in expressions
for the $\Gamma^{(2)}({\bf p},{\bf q})$. Note that we keep the
dependence on the recoil momentum ${\bf\Delta}={\bf
p}_{D_{sJ}^{(*)}}-{\bf p}_{B_s}$.   This replacement removes the relative
momentum dependence in the quark energy   and thus permits us
to perform one of the integrations in the $\Gamma^{(2)}_\mu({\bf p},{\bf q})$
contribution using the quasipotential equation.  The subleading
contribution turns out to be rather small numerically, since it is
proportional to the quark binding energy in the meson. Therefore we
obtain reliable expressions for the form factors in the whole
accessible kinematical range. It is important to emphasize that when doing
these calculations we consistently take into account all relativistic
contributions including boosts of the meson wave functions from the
rest reference frame to the moving ones, given
by Eq.~(\ref{wig}).   The obtained expressions for the decay
form factors are rather cumbersome and are given in the Appendix of Ref.~\cite{bcexc}. 
Note that, while calculating form
factors of weak $B_s$ decays to $D_{s1}$
and $D'_{s1}$  mesons, it is important to take into account the mixing (\ref{eq:mix}) of
singlet $D_s(^1P_1)$ and triplet $D_s(^3P_1)$ states. 

\begin{figure}
  \centering
  \includegraphics[width=7.9cm]{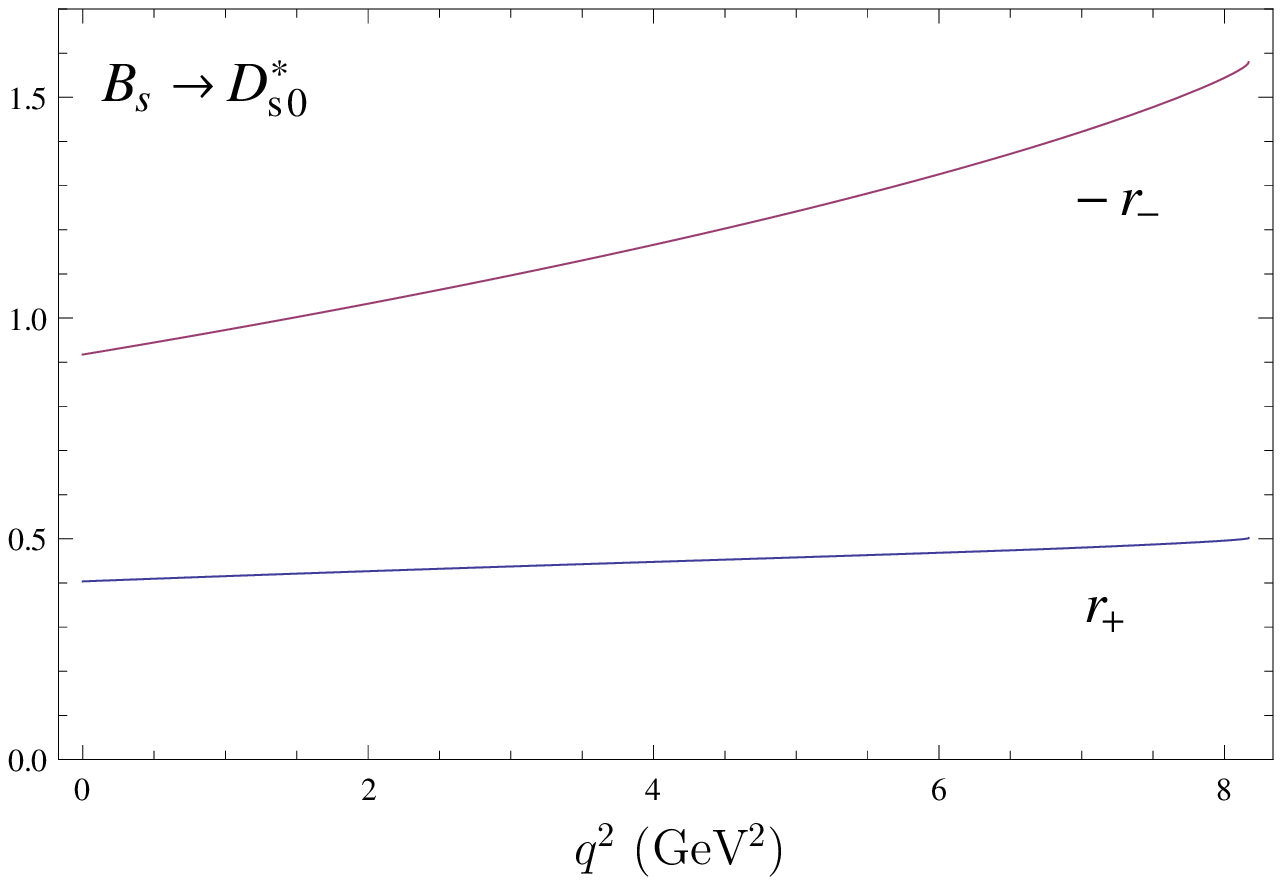} \ \ 
\  
\includegraphics[width=7.9cm]{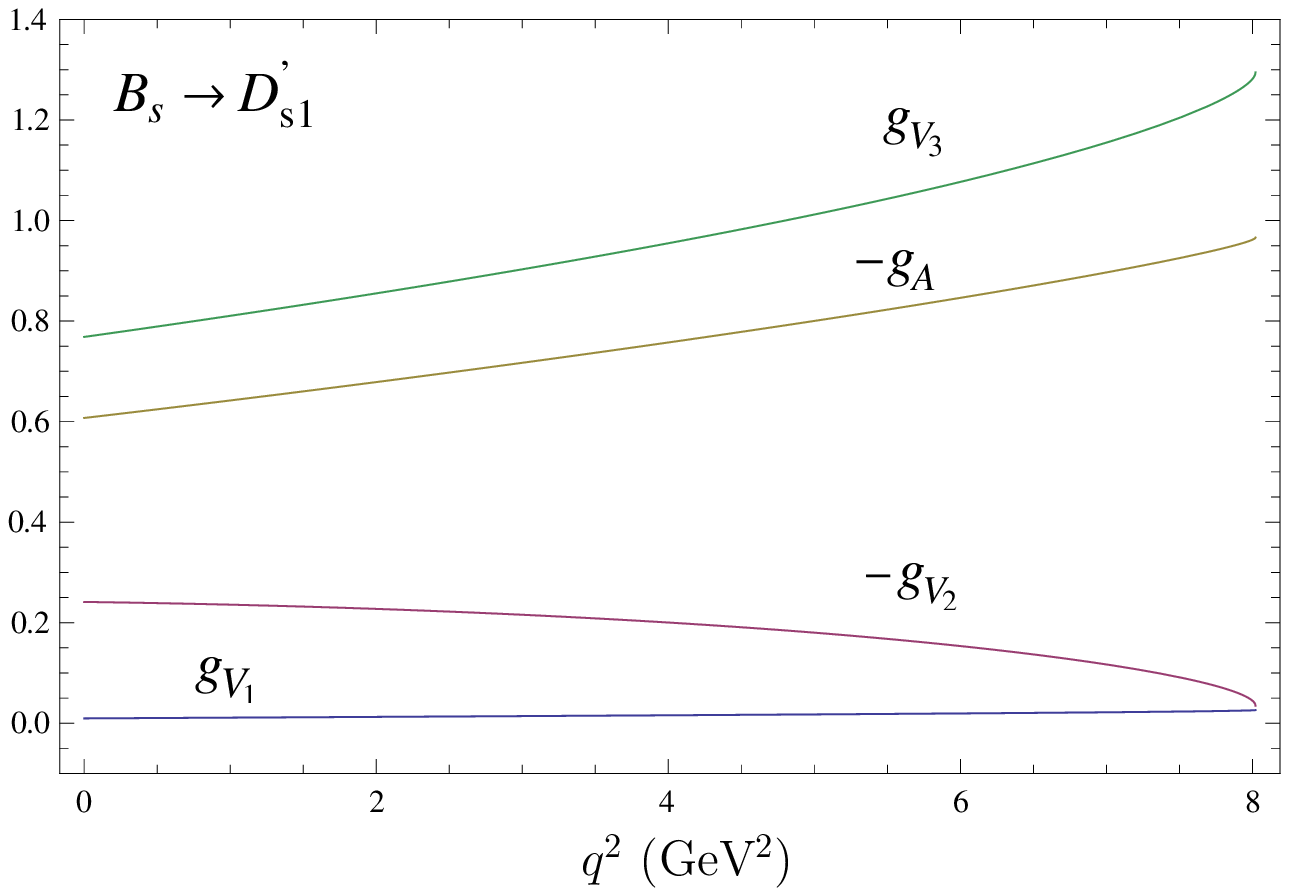}

\vspace*{0.5cm}

  \includegraphics[width=7.9cm]{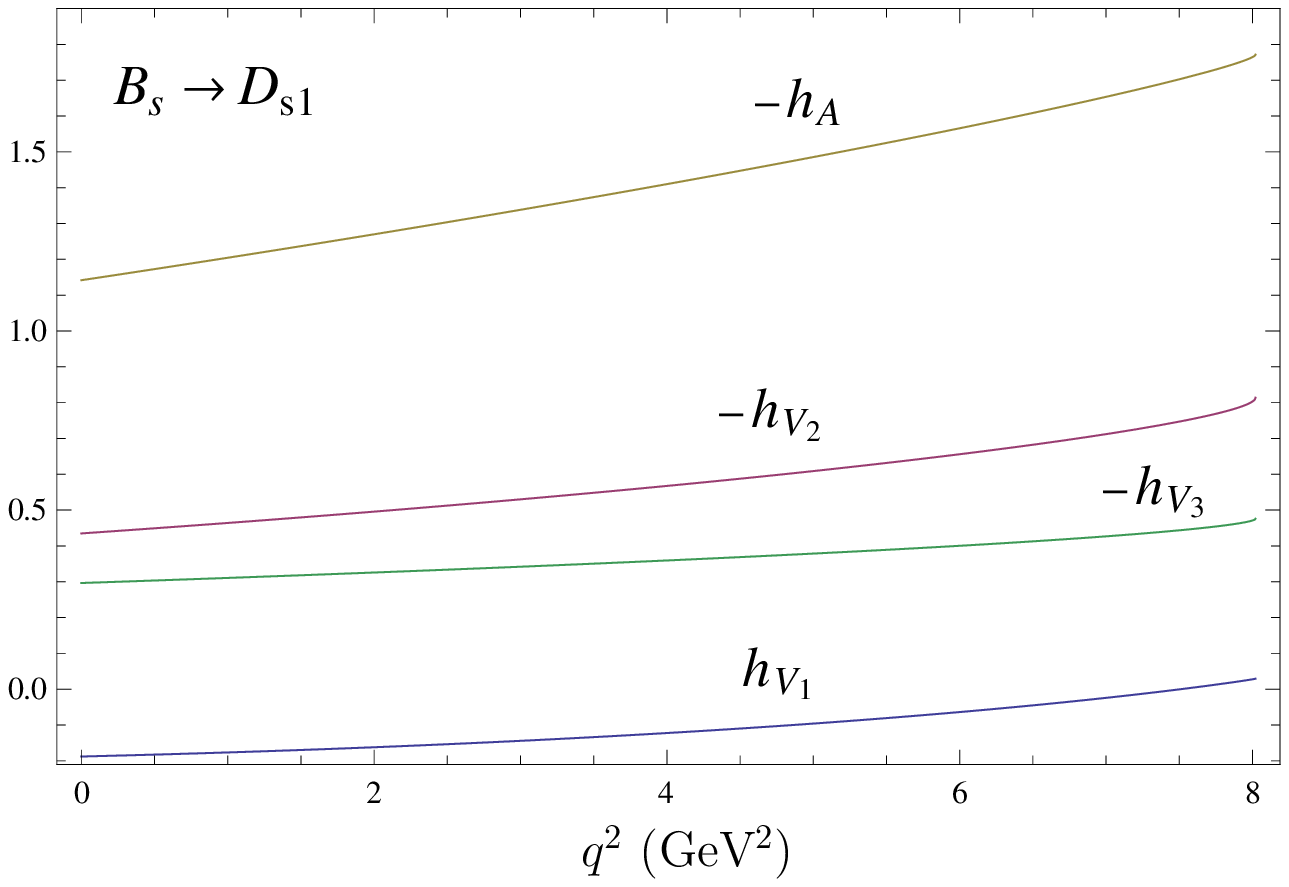} \ \ 
\  
\includegraphics[width=7.9cm]{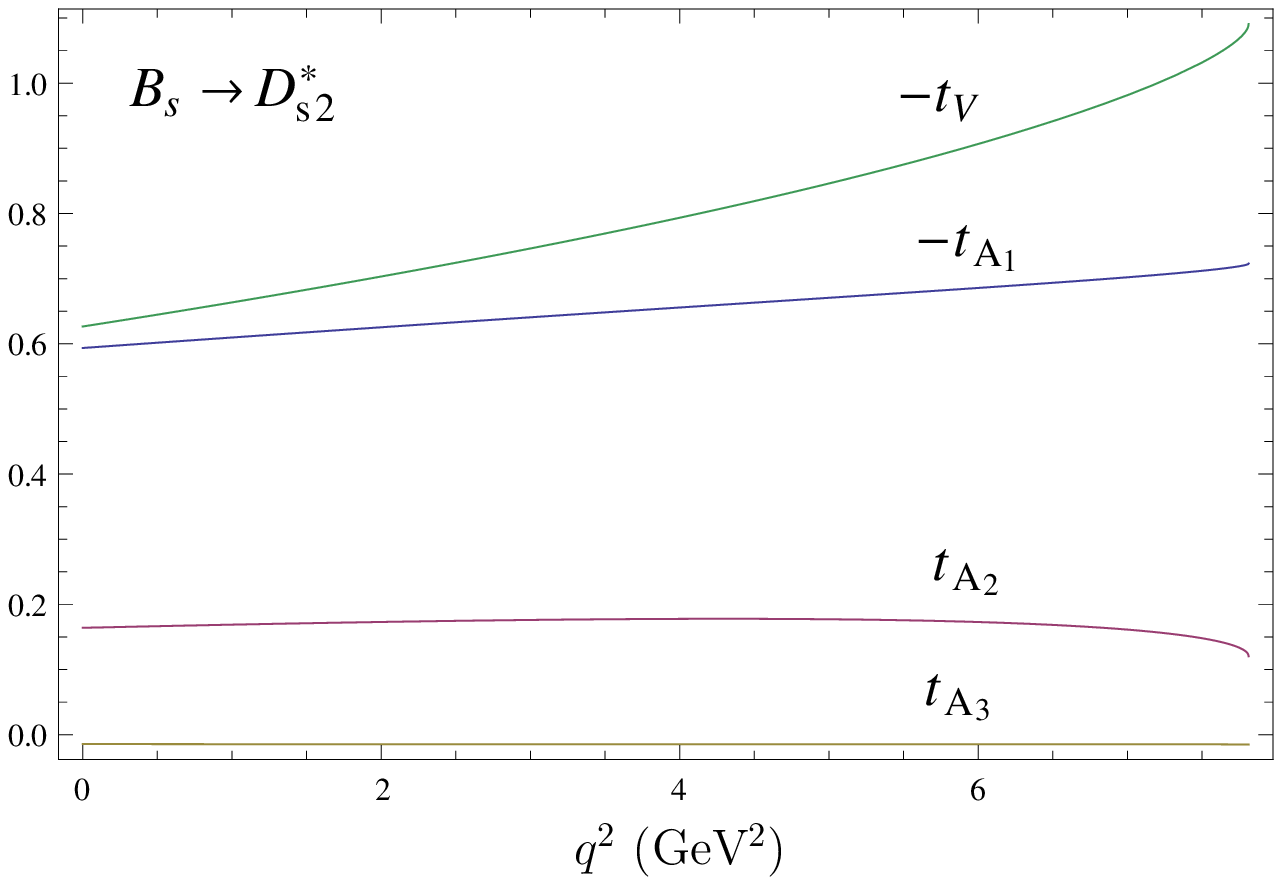}

  \caption{Form factors of the $B_s$ decays to the $P$--wave $D_{sJ}^{(*)}$ mesons.}
  \label{fig:ffbc}
\end{figure}

\begin{table}
\caption{Calculated values of the form factors  of the $B_s$ decays to
  the $P$--wave $D_{sJ}^{(*)}$
  at $q^2=0$ and $q^2=q^2_{\rm max}\equiv(M_{B_s}-M_{D_{sJ}})^2$. }
\label{ffm}
\begin{ruledtabular}
\begin{tabular}{ccccccccccccccc}
&\multicolumn{2}{c}{\underline{\hspace{0.2cm}$B_s\to D_{s0}^*$\hspace{0.2cm}}}&\multicolumn{4}{c}{\underline{\hspace{1.3cm}$B_s\to D_{s1}'$\hspace{1.3cm}}}&\multicolumn{4}{c}{\underline{\hspace{1.3cm}$B_s\to D_{s1}$\hspace{1.3cm}}}&\multicolumn{4}{c}{\underline{\hspace{1.3cm}$B_s\to D_{s2}^*$\hspace{1.3cm}}}\\
$q^2$&$r_+$&$r_-$&$g_A$&$g_{V_1}$&$g_{V_2}$&$g_{V_3}$&$h_A$&$h_{V_1}$&$h_{V_2}$&$h_{V_3}$&$t_V$&$t_{A_1}$&$t_{A_2}$&$t_{A_3}$\\
\hline
0&0.40 &$-0.91$& $-0.61$&$0.01$&$-0.24$&0.77&$-1.14$ &$-0.19$ & $-0.43$&$-0.29$
&$-0.63$&$-0.59$&$0.16$&$-0.01$\\
$q^2_{\rm max}$&0.50 &$-1.58$& $-0.97$&0.03& $-0.04$&$1.29$&$-1.77$ & $0.03$ &$-0.81$ &$-0.48$ 
&$-1.09$& $-0.72$&$0.12$&$-0.02$\\
\end{tabular}
\end{ruledtabular}
\end{table}
In Fig.~\ref{fig:ffbc} we plot form factors of the weak $B_s$ transitions
to the $P$-wave $D_{sJ}^{(*)}$ mesons. The calculated values of these form factors at
$q^2=0$ and $q^2=q^2_{\rm max}\equiv(M_{B_s}-M_{D_{sJ}})^2$ are
displayed in Table~\ref{ffm}. The theoretical uncertainties of these
form factors within our approach are mainly determined by the errors
introduced by the replacement of  $\epsilon_c(p+\Delta)$ by
$\epsilon_c(\Delta)$ in the subleading vertex $\Gamma^{(2)}({\bf
  p},{\bf q})$ and $O(1/m_b^3)$ contributions. They are almost negligible at $q^2=0$ and are less than
1\% at  $q^2=q^2_{\rm max}$.

\section{Semileptonic $B_s$ decays to orbitally excited $D_{sJ}^{(*)}$
  mesons}
\label{sdoe}

The differential semileptonic decay rates of $B_s$ mesons to orbitally
excited $D_{sJ}^{(*)}$ mesons are given by Eq.~(\ref{eq:dgamma}). The
helicity components $H_\pm$, $H_0$ and $H_t$ of the hadronic tensor
are expressed through the invariant form factors
(\ref{eq:sff1})-(\ref{eq:tff2}) by the relations \cite{bcexc} given in
Appendix~\ref{sec:hco}.

\begin{table}
\caption{Comparison of the predictions for the branching fractions of the semileptonic
  decays $B_s\to D_{sJ}^{(*)} l\nu$ (in \%).  }
\label{comphlffoe}
\begin{ruledtabular}
\begin{tabular}{ccccccccc}
Decay& this paper& $m\to\infty$ & with
$1/m_Q$  & \cite{saefhp} &\cite{zll}&\cite{llw}&\cite{as}
&\cite{h}\\
&&\cite{orbexc}&\cite{orbexc}\\
\hline
$B_s\to D_{s0}^*e\nu$&      $0.36\pm0.04$& 0.10 & 0.37 & 0.443 & 0.49-0.571&
$0.23^{+0.12}_{-0.10}$ &$\sim 0.1$ &0.20\\ 
$B_s\to D_{s0}^*\tau\nu$&  $0.019\pm0.002$&   &  &  & & $0.057^{+0.028}_{-0.023}$&$\sim 0.01$
\\ 
$B_s\to D_{s1}'e\nu$&      $0.19\pm0.02$& 0.13 & 0.18 & 0.174-0.570 &
0.752-0.869&   &$ 0.49$& 0.10\\
 $B_s\to D_{s1}'\tau\nu$&      $0.015\pm0.002$  \\ 
$B_s\to D_{s1}e\nu$&      $0.84\pm0.09$& 0.36 & 1.06 & 0.477   \\
 $B_s\to D_{s1}\tau\nu$&    $0.049\pm0.005$&  \\
 $B_s\to D_{s2}^*e\nu$&      $0.67\pm0.07$& 0.56 & 0.75 & 0.376 &  \\ 
 $B_s\to D_{s2}^*\tau\nu$&      $0.029\pm0.003$  \\ 
\end{tabular}
\end{ruledtabular}
\end{table} 

Substituting calculated form factors in these expressions  we get
predictions for the branching fractions of the semileptonic $B_s$
decays to orbitally excited $D_s$ mesons. We find that decays to
$D_{s1}$ and $D^*_{s2}$ mesons are dominant. The obtained results are
given in Table~\ref{comphlffoe} in comparison with other
calculations. First we compare with our previous calculation
\cite{orbexc} which was performed in the framework of the heavy quark
expansion. We give results found in the infinitely heavy quark limit
($m_Q\to\infty$) and with the account of first order $1/m_Q$
corrections. It was argued  \cite{orbexc,llsw} that $1/m_Q$ corrections are large and
their inclusion significantly influences the decays rates. The large
effect of subleading heavy quark corrections was found to be a
consequence of the vanishing of the leading 
order contributions to the decay matrix elements, due to heavy quark
spin-flavour symmetry, at the point of zero recoil of the final charmed
meson, while the subleading order contributions  do not vanish at this
kinematical point. Here we calculated the decay rates without
application of the heavy quark expansion. We find that 
nonperturbative results agree well with the ones
obtained with the account of the leading order $1/m_Q$ corrections \cite{orbexc}. This
means that the  higher order in $1/m_Q$ corrections are small, as was
expected. Then we compare our predictions with the results of
calculations in other approaches. The authors of
Refs.~\cite{saefhp,zll} employ different types of constituent
quark models for their calculations. Light cone and three point QCD sum rules
are used in Refs.~\cite{llw,as}, while HQET and sum rules are applied
in Ref.~\cite{h}. In general we find reasonable
agreement between our predictions and results of
Refs.~\cite{saefhp,llw,as,h}, but results of the quark
model calculations \cite{zll} are slightly larger.

In Fig.~\ref{fig:brboe} we plot  the differential decay
rates  of the $B\to D_{sJ}^{(*)}l\nu$ semileptonic decays. The total semileptonic
decay branching fractions to orbitally excited $D_{s}$ mesons are found
to be  $Br(B_s\to D_{sJ}^{(*)}e\nu)= (2.1\pm0.2)\%$ and $Br(B_s\to D_{sJ}^{(*)}\tau\nu)= (0.11\pm0.01)\%$.

\begin{figure}
  \centering
 \includegraphics[width=8cm]{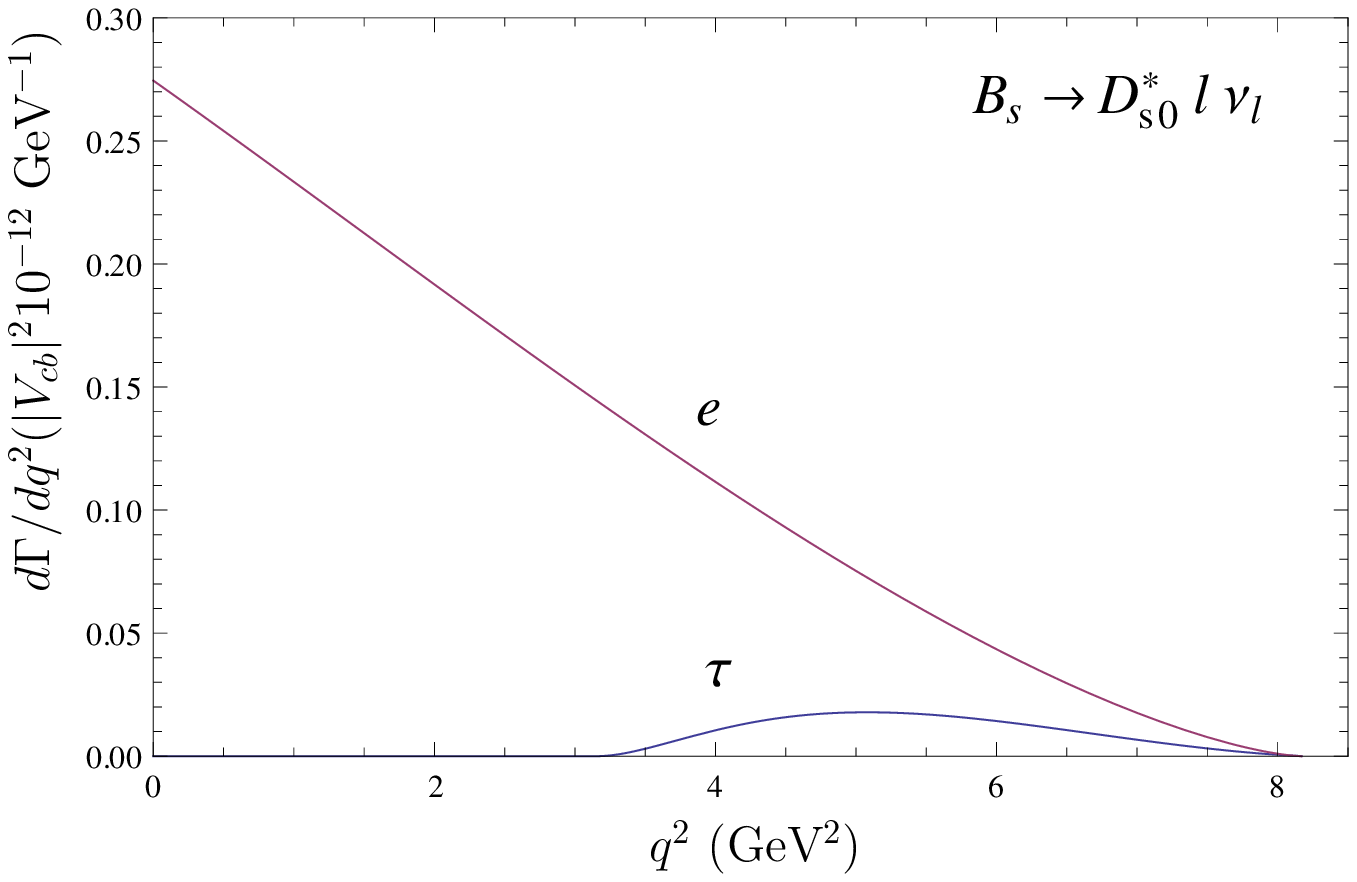}\ \
 \  \includegraphics[width=8cm]{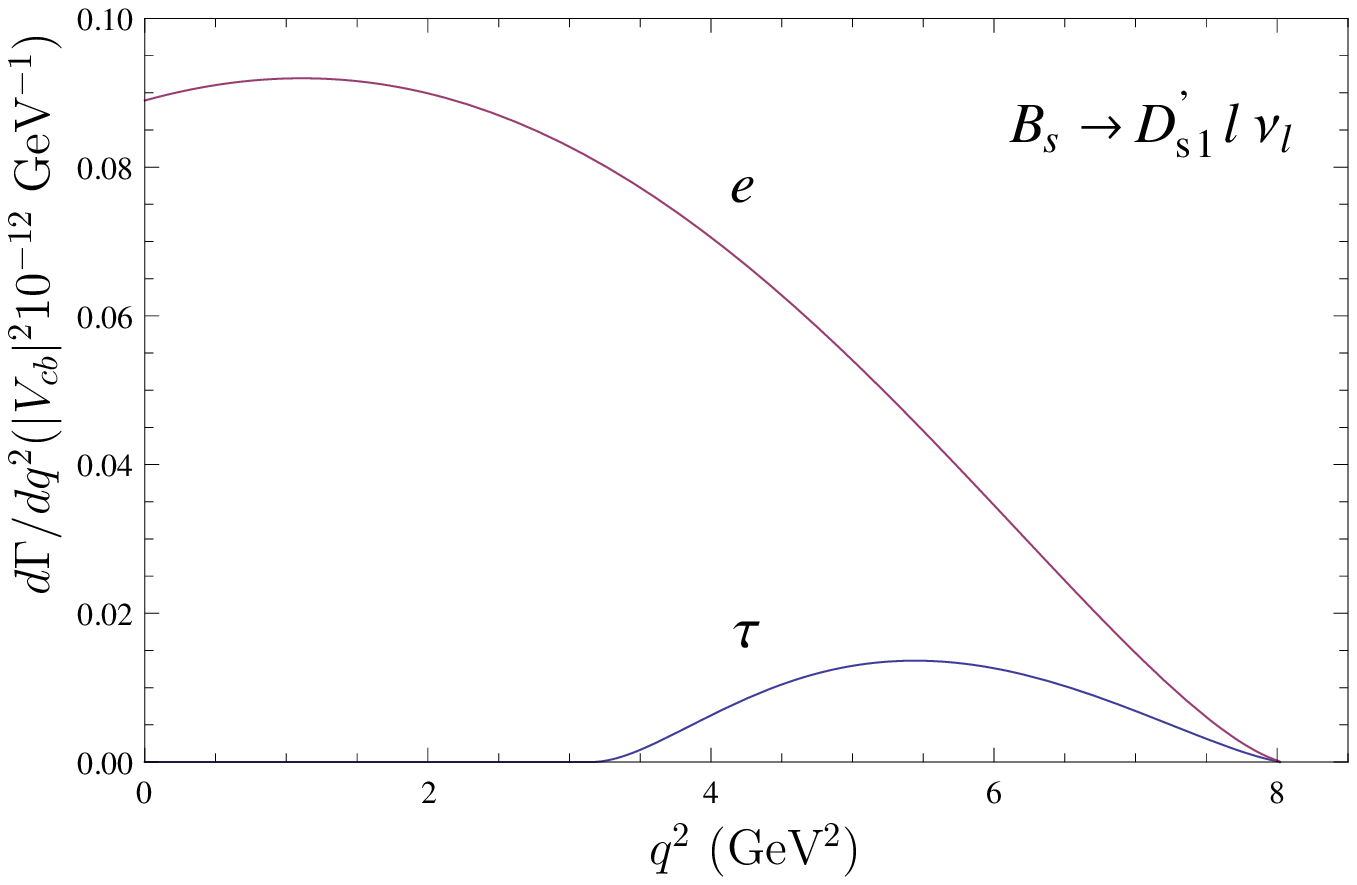}\\
\includegraphics[width=8cm]{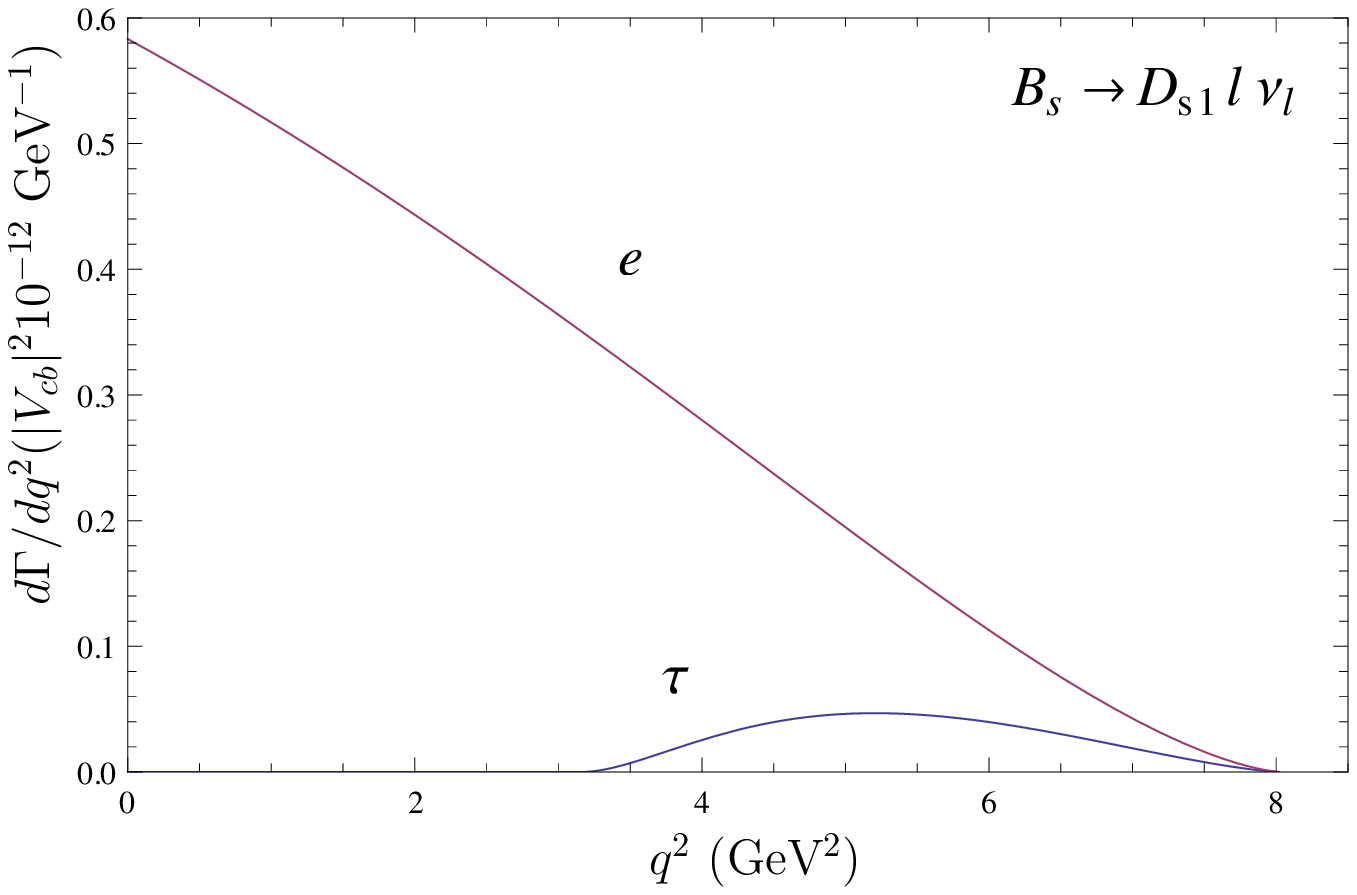}\ \
 \  \includegraphics[width=8cm]{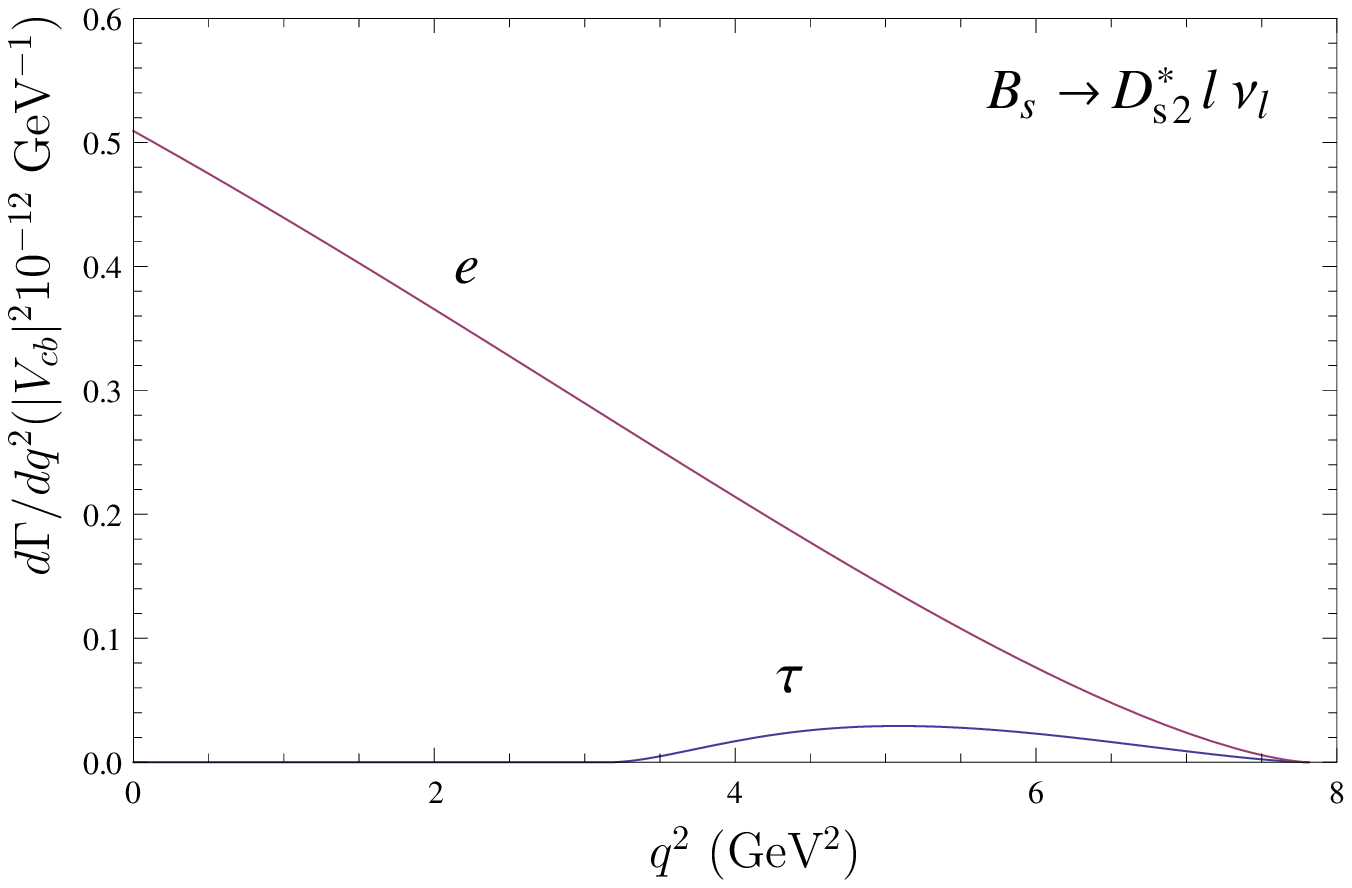}
\caption{Predictions for the differential decay rates  of the $B\to D_{sJ}^{(*)}l\nu$ 
    semileptonic decays. }
  \label{fig:brboe}
\end{figure}

The first experimental measurement of the semileptonic decay $B_s\to
D_{s1}\mu\nu$ was done by the D0 Collaboration \cite{d0}. The
branching fraction was obtained by assuming that the $D_{s1}$ 
production in semileptonic decay comes entirely from the $B_s$ decay  and using
a prediction for $Br(D_{s1}\to D^*K^0_S)=0.25$. Its value $$Br(B_s \to
D_{s1} X\mu\nu)_{\rm D0}=(1.03\pm 0.20\pm 0.17\pm0.14)\%$$ is in good agreement
with our prediction $0.84\pm0.9$ given in Table~\ref{comphlffoe}.

Recently the LHCb Collaboration \cite{lhcb3} reported the first 
observation of the orbitally excited $D_{s2}^*$ meson in the
semileptonic $B_s$ decays. The decay to the $D_{s1}$ meson was also
observed. The measured branching fractions relative 
to the total $B_s$ semileptonic rate are $$Br(B_s\to
D_{s2}^*X\mu\nu)/Br(B_s\to X\mu\nu)_{\rm LHCb} =(3.3\pm1.0\pm0.4)\%,$$   $$Br(B_s\to
D_{s1}X\mu\nu)/Br(B_s\to X\mu\nu)_{\rm LHCb} =(5.4\pm1.2\pm0.5)\%.$$ The
$D_{s2}^*/D_{s1}$ event ratio is found to be $$Br(B_s\to
D_{s2}^*X\mu\nu)/Br(B_s\to D_{s1}X\mu\nu)_{\rm LHCb}=0.61\pm0.14\pm0.05.$$ These
values can be compared with our predictions if we assume that decays
to $D_{s1}$ and $D_{s2}^*$ mesons give dominant contributions to the ratios. Summing up the semileptonic
$B_s$ decay branching fractions to ground state, first radial and
orbital excitations of $D_s$ mesons, presented in Secs.~\ref{sdgs},
\ref{sdre}, \ref{sdoe}, we get for the total $B_s$ semileptonic rate
$Br(B_s\to X\mu\nu)=(10.2\pm 1.0)\%$. Then using the calculated values
from Table~\ref{comphlffoe} we get  $$Br(B_s\to
D_{s2}^*\mu\nu)/Br(B_s\to X\mu\nu)_{\rm theor} =(6.5\pm1.2)\%,$$  $$Br(B_s\to
D_{s1}\mu\nu)/Br(B_s\to X\mu\nu)_{\rm theor} =(8.2\pm1.6)\%,$$ and $$Br(B_s\to
D_{s2}^*\mu\nu)/Br(B_s\to D_{s1}\mu\nu)_{\rm theor}=0.79\pm0.14.$$ 
The predicted central values are larger than 
experimental ones, but the results agree with experiment within $2\sigma$.  

\section{Nonleptonic decays}\label{nl}
In the standard model nonleptonic $B_s$ decays are described by the
effective Hamiltonian, obtained by integrating out the $W$-boson
and top quark. For the nonleptonic $B_s$ decay to the ground state or
excited $D_s$ meson and light meson governed by $b\to c$ transition the
effective Hamiltonian is given by \cite{bbl}
\begin{equation}
\label{heff}
H_{\rm eff}=\frac{G_F}{\sqrt{2}}V_{cb}^*V_{uq}\left[c_1(\mu)O_1^u+
c_2(\mu)O_2^u\right],
\end{equation}
where $q=d,s$. For the nonleptonic  $B_s$ decay to two charmed mesons the
effective Hamiltonian  ($\Delta B=1$) \cite{bbl} reads
\begin{equation}
\label{heff2}
H_{\rm eff}=\frac{G_F}{\sqrt{2}}V_{cb}^*V_{cq}
\sum_{i=1}^{10}c_i(\mu)O_i^c.
\end{equation} 

The Wilson coefficients $c_{i}(\mu)$ are evaluated
perturbatively at the $W$ scale and then are evolved down to the
renormalization scale $\mu\approx m_b$ by the renormalization-group
equations. Functions $O_i^{q'}$ are the local four-quark operators. The
tree level operators have the form
\begin{eqnarray}
\label{o12}
O_1^{q'}&=& (\bar b c)_{V-A}(\bar q' q)_{V-A}, 
\cr 
O_2^{q'}&=&(\bar b_j c_i)_{V-A}(\bar q'_j q_i)_{V-A}, \
\end{eqnarray} 
while the functions  $O_i$ ($i=3,..,10$) are the penguin operators.
The following notations are used
$$(\bar qq')_{V\mp A}=\bar q\gamma_\mu(1\mp\gamma_5)q'.$$
The amplitude of the nonleptonic two-body $B_s$  decay 
 to $D_s$ and light $M$ mesons can be expressed through the
matrix element of the effective 
weak Hamiltonian $H_{\rm eff}$ in the following way
\begin{equation}
\label{mel} 
M(B_s\to D_sM)=\langle D_sM|H_{\rm eff}|B_s\rangle= \frac{G_F}{\sqrt{2}}
\Biggl\{
V_{cb}^*V_{uq}\left[c_1\langle D_sM|O_1^u|B_s\rangle+
c_2\langle D_sM|O_2^u|B_s\rangle\right]\Biggr\}.
\end{equation}

The factorization approach, which is widely used for the calculation
of two-body nonleptonic decays, such as $B_s\to D_sM$, assumes that the
nonleptonic decay amplitude reduces to the product of a meson
transition matrix element
and a weak decay constant \cite{bsw}. Clearly, this assumption is not
exact.  However, it is expected that factorization can hold 
for energetic decays, where one final meson is heavy and the other
meson is light and energetic \cite{dg}. A more general treatment of factorization is given in
Ref.~\cite{bbns}.  

Then the  $B_s\to D_s^-M^+$ decay
amplitude can be approximated by the product of one-particle matrix
elements. The matrix element $(q=d,s)$ is
given by
\begin{equation}
  \label{eq:fact}
  \langle M^+D_s^-|c_1O_1^u+ c_2O_2^u|B_s^0\rangle \approx a_1 
\langle D_s^-|(\bar b c)_{V-A}|B_s^0\rangle \langle M^+|(\bar u
q)_{V-A}|0\rangle,
\end{equation}
where the Wilson coefficients appear in the following
linear combination
\begin{equation}
\label{amu} 
a_1=c_1+\frac{1}{N_c}c_{2} 
\end{equation}
and $N_c$ is the number of colors.  For numerical calculations we use
the values of Wilson coefficients given in Ref.~\cite{wc}.  

The similar expression holds for
$B_s\to D_s^{(*)-}  D_s^{(*)+}$ decays \cite{llw}, namely
\begin{eqnarray}
  \label{eq:factds}
  \langle D_s^+D_s^-|\sum_{i=1}^{10}c_i(\mu)O_i^c|B_s^0\rangle
  &\approx&
  \left(a_1-\frac{V_{tb}^*V_{ts}}{V_{cb}^*V_{cs}}[a_4+a_{10}+r_q(a_6+a_8)]\right)
  \cr
&&\times
\langle D_s^-|(\bar b c)_{V-A}|B_s^0\rangle \langle D_s^+|(\bar 
c s)_{V-A}|0\rangle,
\end{eqnarray}
where the second term in brackets results from the penguin
contributions, which are small numerically. The coefficients
$a_{2n}=c_{2n}+c_{2n-1}/N_c$, and $r_q$ can be found, e.g., in Ref.~\cite{llw}. 

The matrix element of the weak current $J^W_\mu$ between vacuum and a final
pseudoscalar ($P$) or vector ($V$) meson is parametrized by the decay
constants $f_{P,V}$
\begin{equation}
\langle P|\bar q_1 \gamma^\mu\gamma_5 q_2|0\rangle=if_Pp^\mu_P, \qquad
\langle V|\bar q_1\gamma_\mu q_2|0\rangle=\epsilon_\mu M_Vf_V.
\end{equation}
The pseudoscalar $f_P$ and vector $f_V$ decay constants were
calculated within our model in Ref.~\cite{fpconst}. It was shown that
the complete account of relativistic effects is necessary
to get agreement with experiment for decay constants especially of
light mesons. 
We use the following values of the decay constants: $f_\pi=0.131$~GeV,
$f_\rho=0.208$~GeV, $f_K=0.156$~GeV, $f_{K^*}=0.214$~GeV,
$f_{D_s}=0.260$~GeV and $f_{D_s^*}=0.315$~GeV. The relevant CKM
matrix elements are $|V_{ud}|=0.975$,  $|V_{us}|=0.225$,
$|V_{cs}|=0.973$,  $|V_{ts}|=0.0404$,  $|V_{tb}|=0.999$ \cite{pdg}.

The matrix elements of the weak current between the $B_s$ meson and
the final $D_s$ meson  entering in the factorized nonleptonic decay
amplitude (\ref{eq:fact}) are parametrized by the set of the decay form
factors. Using the form factors obtained in Secs.~\ref{sec:ffsdgs},
\ref{sec:ffrexc}, \ref{sec:fforbexc}
we get predictions for the  branching ratios of the 
nonleptonic $B_s$ decays to ground state and excited $D_s$ mesons
and  present them in Tables~\ref{compnl}, \ref{compnlexc} in comparison with
other calculations and available experimental data. We can roughly
estimate the error of our calculations within the adopted
factorization approach to be about 20\%. It originates from both
theoretical uncertainties in the form factors, effective Wilson
coefficients and experimental uncertainties in the values
of the CKM matrix elements (which are dominant), decay constants and meson masses.

\begin{table}
\caption{Comparison of various predictions for the branching fractions of the nonleptonic
  $B_s$  decays to ground state $D_s$ mesons with
  experiment (in $10^{-3}$).  }
\label{compnl}
\begin{ruledtabular}
\begin{tabular}{ccccccccc}
Decay&\hspace*{-1cm}this paper&\cite{bcnp} &\cite{cfkw}&\cite{llw}&\cite{ikkss} & \cite{akf}
&\cite{llz}&Experiment \cite{pdg} \\
\hline
$B_s\to D_s^-\pi^+$& 3.5& 5 & $2.7^{+0.2}_{-0.3}$&$1.7^{+0.7}_{-0.6}$& & $1.42\pm0.57$ & $1.96^{+1.23}_{-0.97}$&
$3.2\pm0.4$  \\ 
$B_s\to D_s^-\rho^+$& 9.4& 13 & $6.4^{+1.2}_{-1.1}$&$4.2^{+1.7}_{-1.4}$& &  & $4.7^{+2.9}_{-2.3}$&
$7.4\pm1.7$  \\
$B_s\to D_s^{*-}\pi^+$& 2.7& 2 & $3.1^{+0.3}_{-0.2}$& && $2.11\pm0.73$ & $1.89^{+1.20}_{-0.93}$&
$2.1\pm0.6$  \\
$B_s\to D_s^{*-}\rho^+$& 8.7& 13 & $9.0^{+1.5}_{-1.5}$&& &  & $5.23^{+3.34}_{-2.56}$&
$10.3\pm2.6$  \\
$B_s\to D_s^-K^+$& 0.28& 0.4 & $0.21^{+0.02}_{-0.02}$&$0.13^{+0.05}_{-0.04}$& & $0.103\pm0.051$ & $0.170^{+0.087}_{-0.066}$& \\ 
$B_s\to D_s^-K^{*+}$& 0.47& 0.6 & $0.38^{+0.05}_{-0.05}$&$0.28^{+0.1}_{-0.8}$& &$0.050\pm0.022$  & $0.281^{+0.147}_{-0.109}$& \\
$B_s\to D_s^{*-}K^+$& 0.21& 0.2 & $0.24^{+0.02}_{-0.02}$& & & $0.159\pm0.067$ & $0.164^{+0.084}_{-0.064}$& \\
$B_s\to D_s^{*-}K^{*+}$& 0.48& 0.6 & $0.56^{+0.06}_{-0.07}$& & &$0.163\pm
0.086$& $0.322^{+0.183}_{-0.124}$&  \\
$B_s\to D_s^-D_s^+$& 11& 10 & $8.3^{+1.0}_{-1.0}$& $35^{+14}_{-12}$&16.5& $2.17\pm0.82$ & &
$5.3\pm0.9$  \\ 
$B_s\to D_s^-D_s^{*+}$& 10& 8 & $8.4^{+1.2}_{-1.2}$ &$33^{+13}_{-11}$&  &$2.62\pm0.93$ &  \\
$B_s\to D_s^{*-}D_s^+$& 6.1& 4 & $7.0^{+1.6}_{-1.5}$ & & &$2.54\pm0.57$ &  \\
$B_s\!\to\! D_s^-D_s^{*+}\!\!+\!D_s^{*-}D_s^+$& 16.1& 12 & $15.4^{+2.0}_{-1.9}$&
&24.0 &$5.16\pm1.10$  & &$12.4\pm2.1$  \\
$B_s\to D_s^{*-}D_s^{*+}$& 25& 16 & $24^{+4}_{-4}$& &31.8 &$27.7\pm7.6$  & &
$18.8\pm3.4$  \\
$B_s\to D_s^{(*)-}D_s^{(*)+}$& 52.1& 38 & $47.7^{+4.6}_{-4.6}$ & &72.3  &$35.0\pm7.8$ &&
$45\pm14$  \\
\end{tabular}
\end{ruledtabular}
\end{table} 

In Table~\ref{compnl} we give predictions for the branching ratios of
the two-body nonleptonic $B_s$ decays to the ground state $D_s^{(*)}$
meson and light ($\pi$, $\rho$, $K^{(*)}$) or heavy $D_s^{(*)}$
meson. We compare our results with predictions of the QCD sum rules
\cite{bcnp}, relativistic constituent quark models \cite{cfkw,ikkss},
the light cone \cite{llw} and 
three-point QCD sum rules \cite{akf}, the perturbative QCD approach
\cite{llz}. Available experimental data \cite{pdg} are also given. We
find reasonable agreement between our results, QCD sum rules \cite{bcnp} and
quark model \cite{cfkw}  predictions and experimental data. Results of
quark model calculation \cite{ikkss} are slightly larger, while those
of three-point QCD sum rules \cite{akf} and perturbative QCD
\cite{llz} are slightly smaller. However, experimental and theoretical
uncertainties are still too large to make possible the discrimination
between theoretical approaches.

\begin{table}
\caption{Branching fractions of the nonleptonic
  $B_s$  decays to orbitally and radially excited  $D_s$ mesons (in $10^{-3}$).  }
\label{compnlexc}
\begin{ruledtabular}
\begin{tabular}{ccc}
Decay& this paper&\cite{llw}\\
\hline
$B_s\to D_{s0}^{*-}\pi^+$& 0.9&  $0.52^{+0.25}_{-0.21}$ \\ 
$B_s\to D_{s0}^{*-}\rho^+$& 2.2& $1.3^{+0.6}_{-0.5}$\\
$B_s\to D_{s0}^{*-}K^+$& 0.07&  $0.04^{+0.02}_{-0.02}$ \\ 
$B_s\to D_{s0}^{*-}K^{*+}$& 0.12& $0.08^{+0.04}_{-0.03}$\\
$B_s\to D_{s0}^{*-}D_s^+$& 1.1& $13^{+7}_{-5}$ \\ 
$B_s\to D_{s0}^{*-}D_s^{*+}$& 2.3&$6.0^{+2.9}_{-2.4}$\\
$B_s\to D_{s1}^{'-}\pi^+$& 0.29&  \\
$B_s\to D_{s1}^{'-}\rho^+$& 0.83&  \\
$B_s\to D_{s1}^{'-}K^+$& 0.021&  \\
$B_s\to D_{s1}^{'-}K^{*+}$& 0.044&  \\
$B_s\to D_{s1}^{'-}D_s^+$& 0.54&  \\
$B_s\to D_{s1}^{'-}D_s^{*+}$& 1.5&  \\
$B_s\to D_{s1}^{-}\pi^+$& 1.9&  \\
$B_s\to D_{s1}^{-}\rho^+$& 4.9&  \\
$B_s\to D_{s1}^{-}K^+$& 0.14&  \\
$B_s\to D_{s1}^{-}K^{*+}$& 0.26&  \\
$B_s\to D_{s1}^{-}D_s^+$& 3.0&  \\
$B_s\to D_{s1}^{-}D_s^{*+}$& 5.9&  \\
$B_s\to D_{s2}^{*-}\pi^+$& 1.6&  \\ 
$B_s\to D_{s2}^{*-}\rho^+$& 4.2&\\
$B_s\to D_{s2}^{*-}K^+$& 0.12& \\ 
$B_s\to D_{s2}^{*-}K^{*+}$& 0.22&\\
$B_s\to D_{s2}^{*-}D_s^+$& 1.4& \\ 
$B_s\to D_{s2}^{*-}D_s^{*+}$& 4.5&\\
$B_s\to D_{s}(2S)^{-}\pi^+$& 0.7&  \\
$B_s\to D_{s}(2S)^{-}\rho^+$& 1.7&  \\
$B_s\to D_{s}^*(2S)^{-}\pi^+$& 0.8&  \\
$B_s\to D_{s}^*(2S)^{-}\rho^+$& 2.2&  \\
$B_s\to D_{s}(2S)^{-}K^+$& 0.05&  \\
$B_s\to D_{s}(2S)^{-}K^{*+}$& 0.08&  \\
$B_s\to D_{s}^*(2S)^{-}K^+$& 0.06&  \\
$B_s\to D_{s}^*(2S)^{-}K^{*+}$& 0.12&  \\
$B_s\to D_{s}(2S)^{-}D_s^+$& 1.0&  \\
$B_s\to D_{s}(2S)^{-}D_s^{*+}$& 0.7&  \\
$B_s\to D_{s}^*(2S)^{-}D_s^+$& 0.7&  \\
$B_s\to D_{s}^*(2S)^{-}D_s^{*+}$& 1.7&  \\
\end{tabular}
\end{ruledtabular}
\end{table} 

In Table~\ref{compnlexc} we present our predictions for the two-body
nonleptonic $B_s$ decays to orbitally and radially excited $D_s$ meson
and light or heavy $D_s$ meson. They are compared with the results of
the light cone sum rules \cite{llw}, which are available only for
decays involving the scalar  $D_{s0}^{*-}$ meson. In general, central
values of our predictions for the decays $B_s\to D_{s0}^{*-}M^+$ (where $M$ is a
light meson)  are slightly larger, but both results are
compatible within errors. On the contrary, for decays $B_s\to
D_{s0}^{*-}D_s^{(*+)}$ our results are significantly lower, especially for the $B_s\to
D_{s0}^{*-}D_s^+$ decay. The same pattern of our predictions and
the light cone sum rules results \cite{llw} holds also for the $B_s$
decays to ground state mesons
(see Table~\ref{compnl}). From Table~\ref{compnlexc} we see that some of the
nonleptonic $B_s$ decays to the excited $D_s$ mesons have branching
fractions comparable with the ones for the decays to the ground state $D_s$
mesons, given in Table~\ref{compnl}. 

Very recently the LHCb
Collaboration announced the first observation of the $B_s\to
D_{s1}\pi$ decay \cite{lhcb4}. Only the relative branching fraction of
this decay was measured. However, this observation indicates that we
can expect the measurement of the nonleptonic $B_s$ decays to excited
$D_s$ mesons in near future. 

\section{Conclusions}
\label{sec:concl}

The weak form factors of the $B_s$ decays to ground state $D_s$ mesons, as well
as to first orbital and radial excitations of $D_s$ mesons were
calculated in the framework of the relativistic quark model based on
the quasipotential approach. The heavy quark expansion was applied for
the calculations of the form factors of the weak $B_s$  decays to
$D_s^{(*)}$ and $D_s^{(*)}(2S)$ mesons. The obtained form factors
satisfy all model independent constraints imposed by heavy quark
symmetry and HQET. The leading and subleading Isgur-Wise functions were
expressed through the overlap integrals of the meson wave
functions. The form factors of weak $B_s$ decays to the orbitally excited
$D_{sJ}^{(*)}$ mesons were calculated, by using previously developed methods
\cite{bcexc}.   All relativistic effects,
including contributions of the intermediate negative-energy states and
transformations of the wave functions to the moving reference frame
were consistently taken into account.    
For the numerical evaluations the relativistic wave
functions of $B_s$ and $D_s$ mesons, obtained as the solutions of
quasipotential equation (\ref{quas}) in Ref.~\cite{hlm}, were used. As a result the weak decay form
factors were determined in the whole accessible kinematical range
without applying any additional parametrizations and extrapolations. 
This significantly reduces theoretical uncertainties of the results.

Using these form factors we considered various
semileptonic $B_s$ decays governed by the $b\to c$ weak
transition. The obtained results were compared with previous
calculations based on constituent quark models, light cone sum rules and
QCD sum rules. The following total semileptonic $B_s$ branching
ratios were found: 
\begin{enumerate}
\item  for decays to ground state $D_s^{(*)}$ mesons  $Br(B_s\to
D_s^{(*)} e\nu)=(7.4\pm 0.7)\%$ and $Br(B_s\to D_s^{(*)}
\tau\nu)=(1.92\pm 0.15)\%$;
\item for decays to radially excited
$D_s^{(*)}(2S)$ mesons  $Br(B_s\to
D_s^{(*)}(2S)e\nu)=(0.65\pm0.06)\%$ and $Br(B_s\to
D_s^{(*)}(2S)\tau\nu)=(0.026\pm0.003)\%$;
\item for decays to orbitally excited $D_{sJ}^{(*)}$ mesons 
$Br(B_s\to D_{sJ}^{(*)}e\nu)= (2.1\pm0.2)\%$ and $Br(B_s\to
D_{sJ}^{(*)}\tau\nu)= (0.11\pm0.01)\%$.
\end{enumerate}
 We see that these
branching fractions significantly decrease with excitation. Therefore,
we can conclude that considered decays give the dominant contribution to the
total semileptonic branching fraction $Br(B_s\to D_s e\nu+{\rm
  anything})$. Summing up these contributions we get the value
$(10.2\pm 1.0)\%$, which agrees well with the experimental value    
$Br(B_s\to D_s e\nu+{\rm anything})_{\rm Exp.}=(7.9\pm2.4)\%$
\cite{pdg}. Note that our predictions for the branching ratios of
semileptonic  decays to orbitally excited states $B_s\to D_{s1}\mu\nu$
and $B_s\to D_{s2}\mu\nu$ are in reasonable agreement with recent data
from the D0 \cite{d0} and LHCb \cite{lhcb3} Collaborations. 

The tree-dominated two-body nonleptonic $B_s$ decays to the ground state
or excited $D_s$ meson and the light or charmed meson were calculated in
the framework of the factorization approximation. This allowed us to
express  the decay matrix elements through the products of the weak
form factors and decay constants. The obtained results were compared
with previous calculations and experimental data, which are mostly
available for the decays involving ground state $D_s$ mesons. Good
agreement of our predictions and data was found. Detailed predictions for
decays involving orbitally $D_{sJ}^{(*)}$  and radially $D_s^{(*)}(2S)$ excited  mesons were
obtained. Some of such decays have branching fractions comparable
with the ones for decays to ground state $D_s$ mesons. The following
decay channels were found to be the most promising: (1) decays to
excited $D_s$ and light mesons $B_s\to
D_{s1}^{-}\rho^+$,  $B_s\to D_{s2}^{*-}\rho^+$, $B_s\to
D_{s0}^{*-}\rho^+$, $B_s\to D_{s}^*(2S)^{-}\rho^+$,    $B_s\to
D_{s1}^{-}\pi^+$,  $B_s\to D_{s}(2S)^{-}\rho^+$, $B_s\to D_{s2}^{*-}\pi^+$; (2) decays to excited
and ground state $D_s$ mesons $B_s\to D_{s1}^{-}D_s^{*+}$, $B_s\to
D_{s2}^{*-}D_s^{*+}$, $B_s\to D_{s1}^{-}D_s^+$. Therefore
there are good reasons to expect that these decays will be measured in the
near future. This expectation is confirmed by the very recent observation
of the $B_s\to D_{s1}\pi$ decay by the LHCb Collaboration \cite{lhcb4}.

\acknowledgements
The authors are grateful to D. Ebert, M.~A.~Ivanov, Z.~Ligeti, V. A. Matveev,
M. M\"uller-Preussker and V. I. Savrin  
for  useful discussions.
This work was supported in part by the {\it Russian
Foundation for Basic Research} under Grant No.12-02-00053-a.

\appendix

\section{HQET expressions for the weak form factors of the $B_s$ decays to ground state $D_s$ mesons}
\label{sec:ffg}

In HQET the weak form factors of the $B_s$ decays to ground state $D_s$
mesons  up to $1/m_Q$ order are expressed as follows \cite{n}
\begin{eqnarray}\label{cffgs}
h_{+}&=&\xi+(\varepsilon_c+\varepsilon_b)\left[2\chi_1-4(w-1)\chi_2+
12\chi_3\right],\\\cr
h_{-}&=&(\varepsilon_c-\varepsilon_b)\left[2\xi_3-\bar\Lambda\xi\right],\\\cr 
h_V&=&\xi+\varepsilon_c\left[2\chi_1-4\chi_3+
\bar\Lambda\xi\right]+\varepsilon_b\left[2\chi_1-4(w-1)\chi_2+
12\chi_3+\bar\Lambda\xi
-2\xi_3\right],\\\cr
h_{A_1}&=&\xi+\varepsilon_c\left[2\chi_1-4\chi_3+
\frac{w-1}{w+1}\bar\Lambda\xi\right]\cr\cr
&&+\varepsilon_b\left[2\chi_1-4(w-1)\chi_2+
12\chi_3+
\frac{w-1}{w+1}\left(\bar\Lambda\xi-2\xi_3\right)\right],\ \ \ \ \ \ \
\\\cr
h_{A_2}&=&\varepsilon_c\left[4\chi_2-\frac2{w+1}\left(
\bar\Lambda\xi+\xi_3\right)\right],\\\cr
h_{A_3}&=&\xi+\varepsilon_c\left[2\chi_1-4\chi_2-
4\chi_3+\frac{w-1}{w+1}\bar\Lambda\xi-\frac2{w+1}\xi_3\right]\cr\cr\label{cffgsl}
&&+\varepsilon_b\left[2\chi_1-4(w-1)\chi_2+
12\chi_3+\bar\Lambda\xi-2\xi_3\right],
\end{eqnarray}
where $\varepsilon_Q=1/(2m_Q)$ and $\bar\Lambda=M-m_Q$.

\section{Relations between two popular sets of form factors}
\label{ffr}

\begin{equation}
  \label{eq:ffiwpl}
  f_+(q^2) = \frac{M_{B_s} + M_{D_s}}{2 \sqrt{M_{B_s} M_{D_s}}}
     h_+\!\!\left(\frac{M_{B_s}^2 + M_{D_s}^2 - q^2}{2 M_{B_s} M_{D_s}}\right) - \frac{M_{B_s} - M_{D_s}}{2 \sqrt{M_{B_s} M_{D_s}}} h_-\!\!\left(\frac{M_{B_s}^2 + M_{D_s}^2 - q^2}{2 M_{B_s} M_{D_s}}\right),
\end{equation}
\begin{eqnarray}
  \label{eq:ffiw0}
  f_0(q^2) =\frac1{2 \sqrt{M_{B_s} M_{D_s}}}&\!\!\Biggl[&\!\!\frac{(M_{B_s} +
      M_{D_s})^2 - q^2}{M_{B_s} + M_{D_s}} h_+\!\!\left(\frac{M_{B_s}^2 +
        M_{D_s}^2 - q^2}{2 M_{B_s} M_{D_s}}\right)\cr\cr
&&-\frac{(M_{B_s} -
      M_{D_s})^2 - q^2}{M_{B_s} - M_{D_s}} h_-\!\!\left(\frac{M_{B_s}^2 + M_{D_s}^2 - q^2}{2 M_{B_s} M_{D_s}}\right)\Biggr],
\end{eqnarray}
\begin{equation}
  \label{eq:ffiwv}
  V(q^2) = \frac{M_{B_s} + M_{D_s^*}}{2 \sqrt{M_{B_s} M_{D_s^*}}} h_V\!\!\left(\frac{M_{B_s}^2 + M_{D_s^*}^2 - q^2}{2 M_{B_s} M_{D_s^*}}\right),
\end{equation}
\begin{equation}
  \label{eq:ffiwa1}
  A_1(q^2)=\frac{(M_{B_s} +
      M_{D_s^*})^2 - q^2}{2 \sqrt{M_{B_s} M_{D_s^*}}(M_{B_s} + M_{D_s*})} h_{A_1}\!\!\left(\frac{M_{B_s}^2 + M_{D_s^*}^2 - q^2}{2 M_{B_s} M_{D_s^*}}\right),
\end{equation}
\begin{equation}
  \label{eq:ffiwa2}
  A_2(q^2)=\frac{M_{B_s} +
      M_{D_s^*}}{2 \sqrt{M_{B_s} M_{D_s^*}}}\Biggl[h_{A_3}\!\!\left(\frac{M_{B_s}^2 + M_{D_s^*}^2 - q^2}{2 M_{B_s} M_{D_s^*}}\right)+\frac{ M_{D_s^*}}{M_{B_s}} h_{A_2}\!\!\left(\frac{M_{B_s}^2 + M_{D_s^*}^2 - q^2}{2 M_{B_s} M_{D_s^*}}\right)\Biggr],
\end{equation}
\begin{eqnarray}
  \label{eq:ffiwa0}
  A_0(q^2)\!\!&=&\!\!\frac{(M_{B_s} +
      M_{D_s^*})^2 - q^2}{4 M_{D_s^*}\sqrt{M_{B_s} M_{D_s^*}}}
    h_{A_1}\!\!\left(\frac{M_{B_s}^2 + M_{D_s^*}^2 - q^2}{2 M_{B_s}
        M_{D_s^*}}\right)\cr\cr
&&\!\!-\frac{M_{B_s}^2 -
      M_{D_s^*}^2}{4 M_{D_s^*}\sqrt{M_{B_s}
        M_{D_s^*}}}\Biggl[h_{A_3}\!\!\left(\frac{M_{B_s}^2 + M_{D_s^*}^2 -
        q^2}{2 M_{B_s} M_{D_s^*}}\right)+\frac{ M_{D_s^*}}{M_{B_s}}
    h_{A_2}\!\!\left(\frac{M_{B_s}^2 + M_{D_s^*}^2 - q^2}{2 M_{B_s}
        M_{D_s^*}}\right)\Biggr]\cr\cr
&&\!\!+\frac{q^2}{4 M_{D_s^*}\sqrt{M_{B_s}
        M_{D_s^*}}}\Biggl[h_{A_3}\!\!\left(\frac{M_{B_s}^2 + M_{D_s^*}^2 -
        q^2}{2 M_{B_s} M_{D_s^*}}\right)-\frac{ M_{D_s^*}}{M_{B_s}}
    h_{A_2}\!\!\left(\frac{M_{B_s}^2 + M_{D_s^*}^2 - q^2}{2 M_{B_s}
        M_{D_s^*}}\right)\Biggr], \ \ \ \ \ \ \ 
\end{eqnarray}

\section{Helicity components of the hadronic tensor for the $B_s\to D_s^{(*)}l\nu$ decays}
  \label{sec:hc}
(a) $B_s\to D_s$ transition
\begin{eqnarray}
  \label{eq:hsa}
  H_\pm&=&0,\cr
H_0&=&\frac{\lambda^{1/2}}{\sqrt{q^2}}f_+(q^2),\cr
H_t&=&\frac1{\sqrt{q^2}}(M_{B_s}^2-M_{D_s}^2)f_0(q^2).
\end{eqnarray}

(b) $B_s\to D_s^*$ transition
\begin{eqnarray}
  \label{eq:helamp}
  H_\pm(q^2)&=&\frac{\lambda^{1/2}}{M_{B_s}+M_{D_s^*}}\left[V(q^2)\mp
\frac{(M_{B_s}+M_{D_s^*})^2}{\lambda^{1/2}}A_1(q^2)\right],\\
  \label{eq:h0a}
  H_0(q^2)&=&\frac1{2M_{D_s^*}\sqrt{q^2}}\left[(M_{B_s}+M_{D_s^*})
(M_{B_s}^2-M_{D_s^*}^2-q^2)A_1(q^2)-\frac{\lambda}{M_{B_s}
+M_{D_s^*}}A_2(q^2)\right], \ \ \ \ \ \ \\
  \label{eq:hta}
  H_t&=&\frac{\lambda^{1/2}}{\sqrt{q^2}}A_0(q^2).
\end{eqnarray}
Here the subscripts $\pm,0,t$ denote transverse, longitudinal and time helicity
components, respectively.

\section{HQET expressions for the weak form factors of the $B_s$ decays
  to radially excited $D_s$ mesons}
\label{sec:ffre}

In HQET the structure of the weak decay form factors   for
$B_s$ decays to radially excited $D_s[(n+1)S]$ mesons  up to
$1/m_Q$ order is the following \cite{radexc}
\begin{eqnarray}\label{cff}
h_{+}&=&\xi^{(n)}+\varepsilon_c\left[2\tilde\chi_1-4(w-1)\tilde\chi_2+
12\tilde\chi_3\right]+\varepsilon_b\chi_b,\\ \cr\label{cff1}
h_{-}&=&\varepsilon_c\left[2\tilde\xi_3-\left(\bar\Lambda^{(n)}+
\frac{\bar\Lambda^{(n)}-\bar\Lambda}{w-1}\right)\xi^{(n)}\right]
- \varepsilon_b\left[2\tilde\xi_3-\left(\bar\Lambda-
\frac{\bar\Lambda^{(n)}-\bar\Lambda}{w-1}\right)\xi^{(n)}\right],\\ \cr\label{cff2}
h_V&=&\xi^{(n)}+\varepsilon_c\left[2\tilde\chi_1-4\tilde\chi_3+
\left(\bar\Lambda^{(n)}+\frac{\bar\Lambda^{(n)}
-\bar\Lambda}{w-1}\right)\xi^{(n)}\right]\cr \cr\label{cff3}
&&+\varepsilon_b\left[\chi_b+
\left(\bar\Lambda-\frac{\bar\Lambda^{(n)}-\bar\Lambda}{w-1}\right)\xi^{(n)}
-2\tilde\xi_3\right],\\  \cr\label{cff4}
h_{A_1}&=&\xi^{(n)}+\varepsilon_c\left[2\tilde\chi_1-4\tilde\chi_3+
\frac{w-1}{w+1}\left(\bar\Lambda^{(n)}+\frac{\bar\Lambda^{(n)}
-\bar\Lambda}{w-1}\right)\xi^{(n)}\right]\cr \cr\label{cff5}
&&+\varepsilon_b\left\{\chi_b+
\frac{w-1}{w+1}\left[
\left(\bar\Lambda-\frac{\bar\Lambda^{(n)}-\bar\Lambda}{w-1}\right)\xi^{(n)}
-2\tilde\xi_3\right]\right\},\\ \cr\label{cff6}
h_{A_2}&=&\varepsilon_c\left\{4\tilde\chi_2-\frac2{w+1}\left[
\left(\bar\Lambda^{(n)}+\frac{\bar\Lambda^{(n)}
-\bar\Lambda}{w-1}\right)\xi^{(n)}+\tilde\xi_3\right]\right\},\\ \cr\label{cff7}
h_{A_3}&=&\xi^{(n)}+\varepsilon_c\left[2\tilde\chi_1-4\tilde\chi_2-
4\tilde\chi_3+\frac{w-1}{w+1}\left(\bar\Lambda^{(n)}+\frac{\bar\Lambda^{(n)}
-\bar\Lambda}{w-1}\right)\xi^{(n)}-\frac2{w+1}\tilde\xi_3\right]\cr\cr\label{cff8}
&&+\varepsilon_b\left[\chi_b+
\left(\bar\Lambda-\frac{\bar\Lambda^{(n)}-\bar\Lambda}{w-1}\right)\xi^{(n)}
-2\tilde\xi_3\right],
\end{eqnarray}
where $\bar\Lambda(\bar\Lambda^{(n)})=M(M^{(n)})-m_Q$ is the difference between
the heavy ground state (radially excited) meson and heavy quark masses in 
the limit $m_Q\to\infty$.

  \section{Helicity components of the hadronic tensor for the $B_s\to D_{sJ}^{(*)}l\nu$ decays}
  \label{sec:hco}

(a) $B_s\to D_{s0}^*$ transition
\begin{eqnarray}
  \label{eq:has}
  H_\pm&=&0,\\
H_0&=&\frac{\lambda^{1/2}}{\sqrt{q^2}}r_+(q^2),\\
H_t&=&\frac1{\sqrt{q^2}}[(M_{B_s}^2-M_{ D_{s0}}^2)r_+(q^2)+q^2r_-(q^2)].
\end{eqnarray}

(b) $B\to D_{s1}$ transition
\begin{eqnarray}
  \label{eq:haav}
  H_\pm&=&(M_{B_s}+M_{ D_{s1}})h_{V_1}(q^2) \pm\frac{\lambda^{1/2}}{M_{B_s}+M_{ D_{s1}}}h_A(q^2),\\
H_0&=&\frac{1}{2M_{ D_{s1}}\sqrt{q^2}} \Biggl\{(M_{B_s}+M_{
    D_{s1}})(M_{B_s}^2-M_{
    D_{s1}}^2-q^2)h_{V_1}(q^2)\cr
&&+\frac{\lambda}{2M_{B_s}}[h_{V_2}(q^2)+h_{V_3}(q^2)]\Biggr\},\
\ \ \ \ \\\label{eq:haavl}
H_t&=&\frac{\lambda^{1/2}}{2M_{ D_{s1}}\sqrt{q^2}}
\Biggl\{(M_{B_s}+M_{ D_{s1}})h_{V_1}(q^2)+\frac{M_{B_s}^2-M_{ D_{s1}}^2}{2M_{B_s}}[h_{V_2}(q^2)+h_{V_3}(q^2)]\cr
&&+\frac{q^2}{2M_{B_s}}[h_{V_2}(q^2)-h_{V_3}(q^2)]\Biggr\}.
\end{eqnarray}

(c) $B\to D'_{s1}$ transition\\
\nopagebreak
\indent $H_i$ are obtained from
expressions (\ref{eq:haav})-(\ref{eq:haavl}) by the replacement of
form factors $h_i(q^2)$ by $g_i(q^2)$ and the final meson mass $M_{
  D_{s1}}$ by $M_{ D'_{s1}}$.

(d) $B\to D_{s2}^*$ transition
\begin{eqnarray}
  \label{eq:haat}
  H_\pm&=&\frac{\lambda^{1/2}}{2\sqrt{2}M_{B_s}M_{D_{s2}}}\left[(M_{B_s}+M_{D_{s2}})t_{A_1}(q^2) \pm\frac{\lambda^{1/2}}{M_{B_s}+M_{D_{s2}}}t_V(q^2)\right],\\
H_0&=&\frac{\lambda^{1/2}}{2\sqrt{6}M_{B_s}M_{D_{s2}}^2\sqrt{q^2}}
\Biggl\{(M_{B_s}+M_{D_{s2}})(M_{B_s}^2-M_{D_{s2}}^2-q^2)t_{A_1}(q^2)\cr
&&+\frac{\lambda}{2M_{B_s}}[t_{A_2}(q^2)+t_{A_3}(q^2)]\Biggr\},\
\ \ \ \ \\
H_t&=&\sqrt{\frac23}\frac{\lambda}{4M_{B_s}M_{D_{s2}}^2\sqrt{q^2}}
\Biggl\{(M_{B_s}+M_{D_{s2}})t_{A_1}(q^2)+\frac{M_{B_s}^2-M_{D_{s2}}^2}{2M_{B_s}}[t_{A_2}(q^2)+t_{A_3}(q^2)]\cr
&&+\frac{q^2}{2M_{B_s}}[t_{A_2}(q^2)-t_{A_3}(q^2)]\Biggr\}.
\end{eqnarray}

\end{document}